\documentclass[aps,prd,twocolumn,showpacs,superscriptaddress,nofootinbib,floatfix]{revtex4-1}  
\usepackage{graphicx}  
\usepackage{dcolumn}   
\usepackage{bm}        
\usepackage{amssymb}   
\usepackage{amsmath}
\usepackage{tabularx}
\usepackage{hyperref}
\usepackage[usenames,dvipsnames]{color}
\usepackage[normalem]{ulem}

\newcommand{\araa}{Annu Rev Astron Astr}
\newcommand{\aap}{Astronomy \& Astrophysics}
\newcommand{\apjl}{The Astrophysical Journal Letters}
\newcommand{\mnras}{MNRAS}
\newcommand{\apjs}{The Astrophysical Journal Supplement Series}
\def\aj{Astronomical Journal}

\newcommand{\msun}{{M}_{\odot}}
\newcommand{\lvsun}{{L}_{V,\odot}}

\newcommand{\rhodmh}{\rho_{\rm DM}}

\newcommand{\rgc}{R}
\newcommand{\rhogcdot}{\dot{\rho}_{\rm GC}}
\newcommand{\rhogc}{\rho_{\rm GC}}
\newcommand{\rhogcn}{\rho_{\rm GC0}}
\newcommand{\ngc}{n_{\rm GC}}
\newcommand{\mspc}{M_{\odot}\,{\rm pc}^{-3}}

\newcommand{\be}{\begin{equation}}
\newcommand{\ee}{\end{equation}}
\newcommand{\bi}{\begin{list}{\labelitemi}{\leftmargin=1em}\setlength{\itemsep}{-3pt}}
\newcommand{\ei}{\end{list}}

\newcommand{\beqn}{\begin{eqnarray}}
\newcommand{\eeqn}{\end{eqnarray}}

\newcommand{\trh}{t_{\rm rh}}
\newcommand{\trhn}{t_{\rm rh,0}}
\newcommand{\rh}{r_{\rm h}}
\newcommand{\rhn}{r_{\rm h,0}}
\newcommand{\rhdot}{\dot{r}_{\rm h}}

\newcommand{\mall}{m_{\rm all}}
\newcommand{\mmean}{\langle \mall\rangle}
\newcommand{\vesc}{v_{\rm esc}}

\newcommand{\kms}{{\rm km\,s}^{-1}}

\newcommand{\mpc}{{\rm Mpc}}
\newcommand{\gpc}{{\rm Gpc}}

\def\dr{{\rm d}}

\def\emcee{{\sc emcee}}
\def\model{{\sc clusterBH}}
\def\modelc{{\sc  BHBdynamics}}

\def\modelb{{\sc clusterBHBdynamics}}
\def\modelbbr{{\sc cBHBd}}

\def\Mc{M_{\rm c}}    
\def\Mcl{M_{\rm cl}}    
\def\Mcldot{\dot{M}}    
\def\Mbh{M_{\rm BH}}   
\def\Mbhdot{\dot{M}_{\rm BH}}   
\def\Mst{M_\star}   
\def\Mstdotsev{\dot{M}_{\star, \rm sev}}   
\def\mlo{m_{\rm lo}}    
\def\Mlo{M_{\rm lo}}    

\def\mmax{m_{\rm max}}    
\def\rhdot{\dot{r}_{\rm h}}
\def\rhdotsev{\dot{r}_{\rm h, sev}}
\def\rhdotrlx{\dot{r}_{\rm h, rlx}}

\def\tcc{t_{\rm cc}}
\def\tsev{t_{\rm sev}}

\def\ntrh{N_{\rm rh}}

\def\Mh{M_{\rm h}}   
\def\Mgc{M_{\rm GCs}}   
\def\ngc{n_{\rm GC}}   
\def\fmerge{K_{\rm merge}}   
\def\nmerge{N_{\rm merge}}   
\def\rh{r_{\rm h}}   
\def\rhn{r_{\rm h,0}}   

\def\phicl{\phi_{\rm cl}}   
\def\phicln{\phi_{\rm cl,0}}   

\usepackage[usenames]{xcolor}

\begin{document}

\preprint{APS/123-QED}

  \title{Merger rate   of black hole binaries from globular clusters: theoretical error bars and comparison to  gravitational wave data {from GWTC-2}
  }
   \author{Fabio Antonini}\thanks{antoninif@cardiff.ac.uk}
    \affiliation{Gravity Exploration Institute, School of Physics and Astronomy, Cardiff University, Cardiff, CF24 3AA, United Kingdom}         
  \author{Mark Gieles}\thanks{mgieles@icc.ub.edu}
  \affiliation{ICREA, Pg. Llu\'{i}s Companys 23, E08010 Barcelona, Spain} 
  \affiliation{Institut de Ci\`{e}ncies del Cosmos (ICCUB), Universitat de Barcelona (IEEC-UB), \\ Mart\'{i} Franqu\`{e}s 1, E08028 Barcelona, Spain}    
\date{\today}

\begin{abstract}
Black hole binaries formed dynamically in globular clusters are believed to be one of the main sources of gravitational waves in the Universe. Here, we use our new population synthesis code, \modelbbr, to determine the redshift evolution of the merger rate density and masses of black hole binaries formed in globular clusters. 
We simulate $\sim 2$ million models to explore the parameter space that is relevant to real globular clusters and over all mass scales.
We show that when uncertainties on the initial cluster mass function and their { initial half-mass density} are properly taken into account, they become the two dominant factors in setting the theoretical error bars on  merger rates. Uncertainties in other model parameters (e.g., natal kicks, black hole masses, metallicity) have virtually no effect on the local merger rate density, although they affect the masses of the merging  black holes.
 Modeling   
 the merger rate density as a function
of redshift as $\mathcal{R}(z)=\mathcal{R}_0\left(1+z \right)^{\kappa}\ $
 at  $z<2$, and marginalizing over uncertainties, we find:
$\mathcal{R}_0=7.2^{+21.5}_{-5.5}{ \rm Gpc^{-3}yr^{-1}} {\ \rm and\ } \kappa=1.6^{+0.4}_{-0.6}\ \nonumber$
($90\%$ credibility).
The rate parameters 
for binaries that merge inside the clusters
are $\mathcal{R}_{\rm 0,in}=1.6^{+1.9}_{-1.0}\,{ \rm Gpc^{-3}yr^{-1}}$ and $\kappa_{\rm in}=2.3^{+1.3}_{-1.0}$; $\sim 20\%$ of these  form as the result of a gravitational-wave capture, implying that
eccentric mergers from globular clusters contribute  
$\lesssim 0.4 \rm \,Gpc^{-3}yr^{-1}$ to the local  rate.
A comparison to the merger rate reported by LIGO-Virgo shows that  a scenario in which most of the detected  black hole mergers are formed in globular clusters is consistent with current constraints, and
requires initial cluster half-mass densities $\gtrsim 10^4\,\mspc$. Interestingly, these models also
reproduce  the inferred  primary black hole mass distribution  in the range $13-30\,\msun$. However, all models under-predict the data outside this range, suggesting that other mechanisms might be responsible for the formation of these sources.
\end{abstract} 
\maketitle
%
%

\section{Introduction}\label{intro}

Several black hole (BH) binaries
have  been detected by the Advanced LIGO and Virgo interferometers \cite{2015CQGra..32g4001L,2015CQGra..32b4001A,2016PhRvL.116f1102A,2016PhRvX...6d1015A,2019ApJ...882L..24A,2019PhRvX...9c1040A,2019PhRvD.100f4064A,2020ApJ...896L..44A,2020arXiv200408342T,2020PhRvL.125j1102A}.
{  The recently released Second  Gravitational-Wave Transient Catalog (GWTC-2),  includes a total of  44 confident BH binary
 (BHB) events \cite{Abbott:2020niy,Abbott:2020gyp}.
}
While the astrophysical origin of these sources is still unknown,
one widely discussed possibility is that they
formed in the dense core of globular clusters (GCs) through dynamical three-body interactions \cite{Sigurdsson1993,Kulkarni1993a,2000ApJ...528L..17P,Banerjee2010,Downing2011}.

The realistic modeling of the dynamical evolution of BHs in the core of a GC represents
a complex computational challenge requiring an enormous dynamical range in
both space and time. For this reason, it is only very recently, thanks to major
improvements in computational methods and hardware, that it became possible
to make robust predictions about the numbers and
physical properties of  BH binary (BHB) mergers produced in GCs \citep{Rodriguez2015a,2017MNRAS.464L..36A,2017MNRAS.469.4665P,2018ApJ...855..124S,2019ApJ...871...91Z}. 
Thanks to these past efforts it is now  clear that a large fraction
of the sources detected by {LIGO-Virgo} could have been
dynamically assembled in GCs. However, as discussed below,
a full characterisation of the model uncertainties related to the BHB merger rate from the GC channel  is still missing.

Several  recent studies  only considered the contribution
from clusters that have survived to the present day \cite[e.g.,][]{2017MNRAS.464L..36A,2020ApJS..247...48K}.
These studies found that the present-day population of GCs
produces BHB mergers at a local rate of $\approx 5 \rm~ Gpc^{-3}\ yr^{-1}$.
This represents a lower limit to the actual merger rate
as there likely existed a population of  clusters which did not survive to the present,
but that contributed significantly to the local merger rate \cite{2018PhRvL.121p1103F}.
In fact, it is believed that the GC  mass function (GCMF) today is the result of
an initial GCMF that was shaped by dynamical processes \cite[e.g.,][]{1997ApJ...474..223G,1997MNRAS.289..898V,2001ApJ...561..751F,Baumgardt2003,2008MNRAS.389L..28G,2007ApJS..171..101J}.
These processes, e.g., relaxation {driven evaporation} and tidal shocking,
are particularly efficient at destroying low-mass clusters.
A key uncertainty in estimating a  merger rate  from \emph{all} GCs
is that the amount of such disrupted clusters is not known.

Previous estimates for the
BHB merger rate ignored the fact that
the  fractional mass that
has been lost from {the GC population} by the present time, $K$ (see equation \ref{fml} below), is very uncertain as
it cannot be
tightly constrained from the present-day properties of the GC population.
We will show that once this uncertainty is taken properly into account, it becomes
one of the dominant factors in setting the error bars on local merger rate estimates from the GC channel.
For example, both \citet{2018PhRvL.121p1103F} and \citet{2018ApJ...866L...5R} used a single value 
for $K$ which was derived under one assumption for the initial GCMF.
Although \citet{2018ApJ...866L...5R} considered the effect of different
GCMFs on their results,
they neglected  that $K$ should be related to the choice of initial GCMF { and that its value can be constrained by the present-day GCMF once evaporation mass loss is taken into account}.
A different initial GCMF not only changes the mass of the GCs which make the BHBs, but
it also sets the amount of mass that is lost from the GC system;
and it turns out that the BHB merger rate is quite sensitive to both effects. 

{T}he  properties and merger rate of BHBs
 depend on several other physical
processes, many of which lack strong observational constraints \cite{2017ApJ...834...68C}.
 For example, the  distribution of the natal kicks
controls the number of BHs that are ejected from the GC upon formation, as well
as the fraction of BHs that retain their binary companion after supernova. 
Different assumptions about the early stages of BH formation will also reflect on the
evolution of the host cluster, affecting its total lifetime and final properties.
Moreover, merger rates
are expected to be sensitive to the assumed density {and the related mass-radius relation} of GCs at formation which
is also unconstrained observationally  \cite{2010MNRAS.408L..16G}.
The full implications of these  uncertainties is still 
not fully explored.
The main reason for this is that standard numerical techniques such as  $N$-body and  Monte Carlo
simulations are 
still too slow to allow a full parameter space exploration.
This is why in this study we employ our new population synthesis code \modelb\ (hereafter \modelbbr) \citep{2020MNRAS.492.2936A0}
to  systematically vary  assumptions made for the model parameters and over the full range of initial conditions
relevant to real GCs. Thus, we examine the effect of these initial assumptions on the number and properties of merging BHBs using a suite of about 20 million cluster models.

In summary,  the  merger rate of BHBs
produced dynamically in GCs has been studied by
multiple teams \citep[e.g.,][]{Banerjee2010,2016PhRvD..93h4029R,2017MNRAS.464L..36A,2017MNRAS.469.4665P}. 
Here we build on  former studies  in two ways which allow us to place  error bars
on theoretical estimates for the BHB merger rate density and on its redshift evolution:
(i) we constrain the fractional mass that has been lost from GCs over cosmic time by
fitting an evolved Schechter mass function to the observed GCMF in the Milky Way today,
 and using a simple model for cluster evaporation.
(ii) we employ our new population synthesis code \modelbbr\ to explore how the BHB merger rate depends on 
uncertain parameters in the models (e.g., initial cluster densities,
BH formation recipes, natal kicks), and explore the  parameter space that is relevant to real GCs and over all mass scales.

The paper is organized as follows. In Section~\ref{GCFR} 
we  compute the GC formation rate density as a function of time using constraints from the present-day GCMF.
In Section~\ref{clev} we describe our population synthesis model
and detail the modifications we made to it with respect to the version used in \citep{2020MNRAS.492.2936A0}.
Section~\ref{MR} describes our main results. 
 We discuss the implications of our results and conclude  in Section~\ref{sec:discussion}.

\section{Cluster formation rate}\label{GCFR}
In order to  compute a BHB merger rate we need the cluster formation rate density (i.e. per unit of volume) as a function of time: $\rhogcdot(t)$.
We do this by imposing that in our model: (i) the present-day GC mass density in the Universe,  $\rhogc$,  is consistent with its empirically inferred value and (ii) the present-day GCMF is consistent with the observed mass function of the Milky Way GCs.
\subsection{Globular clusters density in the Universe}
To derive $\rhogc$ we use the same approach as \cite{Rodriguez2015a}, who use the empirically established relation between the total mass of a GC population ($\Mgc$) and the dark matter halo mass of the host galaxy ($\Mh$).  The ratio of these two quantities is remarkable constant over a large range of halo masses  ($10^{10}\lesssim \Mh/\msun \lesssim10^{15}$) and for different galaxy types: $\eta \equiv \Mgc/\Mh\simeq(3-7)\times10^{-5}$ \cite{2009MNRAS.392L...1S,2010MNRAS.406.1967G,Harris2013, 2015ApJ...806...36H,2017ApJ...836...67H}.  We can also estimate this ratio for the Milky Way: the total luminosity of all Milky Way GCs from the Harris catalogue \cite{1996AJ....112.1487H,2010arXiv1012.3224H} is $1.75\times10^7~\lvsun$. Adopting a mass-to-light ratio in the $V$-band of $\Upsilon_V=2~\msun/\lvsun$ and a virial mass of the Milky Way of $\Mh=1.3\times10^{12}~\msun$ \cite{2019ApJ...873..118W} we find $\eta=2.7\times10^{-5}$ for our Galaxy. Table~1 in \cite{2015ApJ...806...36H} summarises 8 results from different studies. We use  the 7 studies that include at least 25 galaxies and combine this with the result of $\eta=2.9\times10^{-5}$ by \cite{2017ApJ...836...67H} that was published after this summary. We also add the Milky Way estimate from above to find a mean value of

\begin{equation}
\langle\eta\rangle =(4.4\pm1.6)\times10^{-5} \ .
\label{eq:eta}
\end{equation}

We  determine the dark matter halo mass function from simulations of large scale structure formation by \cite{Tinker2010} using the \texttt{HMFcalc} tool \cite{2013A&C.....3...23M}. The total mass density in dark matter halos with individual masses  $\Mh\ge10^{10}~\msun/h$ is $\rhodmh = 3.64\times10^{19}~h^2\,\msun\,{\rm Gpc}^{-3}$. 
Combined with our value for $\eta$ from equation~(\ref{eq:eta}) and $h= 0.674$ from the  Planck Collaboration \cite{2018arXiv180706209P} we  find 

\begin{equation}
\rhogc=\langle\eta\rangle\rhodmh=(7.3\pm2.6)\times 10^{14} {\msun\,{\rm Gpc^{-3}}}. 
\label{eq:rhogc}
\end{equation}

The relation between $\Mgc$ and $\Mh$ may hold down to dwarf galaxy masses of $\Mh\simeq10^9~\msun$ \cite{2018MNRAS.481.5592F}, and including these low-mass galaxies would increase $\rhodmh$ and   $\rhogc$ by about 15\%, but because of the uncertain GC occupation fraction below $\Mh\simeq10^{10}~\msun$, we continue with the result of equation~(\ref{eq:rhogc})\footnote{For an average GC mass of $\langle M\rangle = 3\times10^5~\msun$ this mass density implies a number density of $\ngc=2.4\pm0.9~\mpc^{-3}$ and in section~\ref{sec:discussion} we discuss how this compares to other studies.}.

\subsection{Globular cluster mass function}\label{GCMF}
For our population synthesis model of the next section, we need the initial mass density of GCs in the Universe ($\rhogcn$). 
To find the relation between  $\rhogcn$ and $\rhogc$ from the previous section, we  adopt a simple model for the mass evolution of GCs. 
We  assume that the initial GCMF is a  Schechter-type function \cite{Schechter1976}, i.e. a power-law with index $-2$ at low masses with an exponential high-mass  truncation at $\Mc$, as is found for young massive clusters in the Local Universe \cite{PortegiesZwart2010}. We assume that this initial GCMF is universal throughout the Universe and across cosmic time.
There are arguments for a flatter initial GCMF (i.e. fewer low-mass GCs) in dwarf galaxies at high redshift and low metallicity \cite{2002ApJ...566L...1B}, but we proceed with the assumption of a Universal initial GCMF.
We will discuss the effect of flatter initial GCMFs on the BHB merger rate in section~\ref{sec:discussion}. 

To find an expression for the GCMF today, resulting from the initial GCMF we follow a similar approach as \cite{2001ApJ...561..751F,2007ApJS..171..101J} and
 assume that all GCs have lost an amount of mass $\Delta=|\dot{M}| t$, where $\dot{M}$ is the mass loss rate from escaping stars and $t$ is the age of the GCs. We do not specify the escape mechanism and let $\Delta$ be constrained by the Milky Way GCMF. Details of the various processes can be found in literature: relaxation driven evaporation \cite{2003MNRAS.340..227B};  disc and bulge shocks \cite{1972ApJ...176L..51O, 1999ApJ...514..109G}; interactions with molecular gas clouds (at young ages) \cite{1958ApJ...127...17S, 2006MNRAS.371..793G, 2010ApJ...712L.184E} and  combinations of the various effects \cite{1997ApJ...474..223G, 2001ApJ...561..751F,Kruijssen2015, 2016MNRAS.463L.103G}. {From here on we refer to the mechanism responsible for $\dot{M}$, regardless of what the underlying physical process may be,  as `evaporation'. }
 
We then assume that $\Delta$ is a constant, i.e. independent of GC mass, host galaxy, orbit and formation epoch. This is clearly not realistic, because  $\dot{M}$ depends on the (time-dependent) tidal field and the GC orbit within their galaxy \cite{2003MNRAS.340..227B}. However, this exercise is merely meant to  arrive at an order of magnitude estimate of how much mass GCs lose between formation and now, rather than developing a realistic description of GC evolution. 
The present-day GCMF, $\phicl$,  defined as the number of GCs per unit volume ($\ngc$) in the mass range $[M, M+\dr M]$,  is given by the `evolved Schechter function' \cite{2007ApJS..171..101J}
\begin{equation}
\phicl = A(M+\Delta)^{-2}\exp\left(-\frac{M+\Delta}{\Mc}\right) \ .
\label{eq:evsch}
\end{equation}

At low masses, where the GCMF is affected by mass loss ($M\lesssim \Delta\lesssim\Mc$), this function approaches a constant $\phicl \simeq 
A/\Delta^2$. In fact, any initial GCMF evolves towards a uniform $\phicl$ at low masses  if $\dot{M}$ is constant \cite{1961AnAp...24..369H}. The GCMF is often plotted as the number of GCs in logarithmic mass bins ($\propto\dr N/\dr \log M$), which increases linearly with $M$ at low masses and peaks at $M_{\rm peak}\simeq \Delta$ (for $\Delta\lesssim\Mc$).
The simple functional form of equation~(\ref{eq:evsch}) provides a good description for the Milky Way GCMF and the luminosity function of GCs in external galaxies \cite{2007ApJS..171..101J}. 
The constant of proportionality $A$ is found from the constraint that all GCs must add up to the present-day GC mass density in the Universe: $\int_{\Mlo}^{\infty} \phicl M\dr M =\rhogc$, with $\rhogc$ from equation~(\ref{eq:rhogc}) and  $\Mlo=100~\msun$. 

The GC evolution model we use in the next section also considers mass loss by stellar evolution, which mostly happens in the first few 100 Myr. The fraction of mass that clusters lose as a result of stellar evolution depends on metallicity, the stellar initial mass function,  stellar evolution details and on whether BHs are ejected, or not. 
{ For the cluster evolution model of the next section we need the remaining mass in stars and white dwarfs ($\Mst$). We use SSE to compute $\Mst$ at 11 Gyr for a Kroupa IMF in the range $0.1-100\,\msun$. We find that for metallicities of [0.01, 0.1, 1] Solar, the remaining mass fraction is $\Mst({\rm 11\,Gyr})/\Mst(0) \simeq [0.54, 0.53, 0.55]$. The absence of an obvious metallicity trend, is because the remaining mass fraction of stars(white dwarfs) decreases(increases) with metallicity, in approximately similar magnitudes.}
{ This justifies the assumption that the remaining $\Mst$} is independent of metallicity and when excluding mass loss by BH ejections a cluster loses approximately half of its initial mass by stellar evolution. 
We therefore assume that clusters lose half their mass by stellar evolution alone. Next, we assume that stellar evolution and escape affect the GCMF sequentially (i.e. first stellar mass loss and then escape).  We can then write $M_0 = 2(M+\Delta)$ and we can find the initial GCMF from the continuity equation \cite{2001ApJ...561..751F}

\begin{equation}
\phicln\equiv \frac{\dr \ngc}{\dr M_0} = \phicl(M_0)\left|\frac{\partial M}{\partial M_0}\right|.
\label{eq:gcmf}
\end{equation}
Because  $\partial M/\partial M_0=0.5$ and $\phicl(M_0)=A(M_0/2)^{-2}\exp\left[-M_0/(2\Mc)\right]$,  the initial GCMF that corresponds to the present-day GCMF of equation~(\ref{eq:evsch}) is given by

\begin{equation}
  \phicln = 2A M_0^{-2}\exp\left(-\frac{M_0}{2\Mc}\right) .
\label{CIMF}
\end{equation}
We note that the Schechter mass of the initial GCMF is $2\Mc$, where $\Mc$ is derived from the present-day GCMF. 

We then introduce a factor $K$ for the ratio $\rhogcn$ over $\rhogc$, i.e. 

\begin{equation}\label{fml}
K=  \frac{\rhogcn}{\rhogc} = \frac{\int_{\Mlo}^{\infty}\phicln M_0\dr M_0}{\int_{\Mlo}^{\infty} \phicl M\dr M} \ .
\end{equation}
 
\begin{figure}
    \centering
    \includegraphics[width=3.4in,angle=0.]{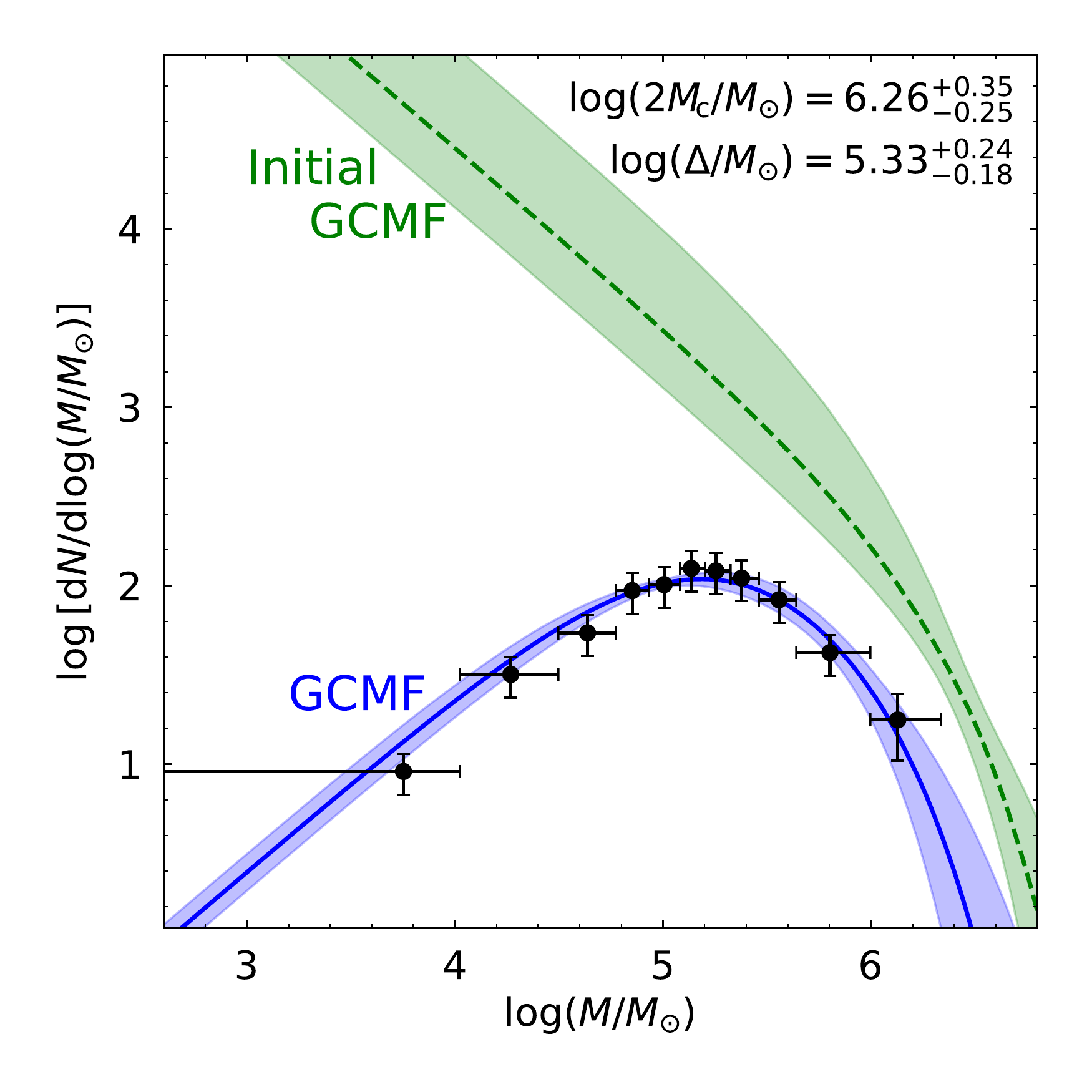}
    \caption{GCMF of 156 Milky Way GCs from the Harris catalogue (black dots with error bars). The blue line shows an evolved Schechter function fit (equation~\ref{eq:evsch}). The resulting initial GCMF, corrected for mass loss by stellar evolution (factor of 2) and evaporation ($\Delta$), is shown as a  green dashed line. The shaded regions and uncertainties of quoted values indicate the {90\% credible} intervals. The inferred $K$ value implies that the total  mass of the Milky Way GC population was 32.5 times higher initially. Half of this mass reduction is because of stellar mass loss and the remaining factor of 16 is due to evaporation.
        } 
    \label{fig1}
\end{figure}

In the next section we include the contribution from clusters of all masses, and it is therefore important to understand the exact value of $K$, or better the distribution of $K$. To find a posterior distribution for $K$, we fit the evolved Schechter functions from equation~(\ref{eq:evsch}) to the Milky Way GCs and then derive $K$ using equations~ (\ref{CIMF}) and (\ref{fml}). We use the $V$-band luminosities of the 156 GCs in the 2010 edition of the Harris catalogue \cite{1996AJ....112.1487H,2010arXiv1012.3224H} and then assume a constant mass-to-light ratio of $\Upsilon_V=2~\msun/\lvsun$ to convert luminosities to masses. A histogram of the resulting mass function is shown in Fig.~\ref{fig1}. The Milky Way values are binned in bins with 15 GCs each, with the highest mass bin containing 6 GCs. The black dots are the average masses of the GCs in each bin, while the horizontal error bars show the bin range and the vertical error bars show the Poisson errors.

We then use the normalised evolved Schechter function of equation~(\ref{eq:evsch}) as a likelihood function to find $\Delta$ and $\Mc$. 
We use the Markov Chain Monte Carlo (MCMC) code \emcee\ \cite{Foreman-Mackey2012}  to maximise the log-likelihood and vary $\log\Delta$ and $\log\Mc$, assuming flat priors in the range $3\le\log(\Delta/\msun)\le7$ and $3\le\log(\Mc/\msun)\le7$. In Fig.~\ref{fig1} we show the result.  For $10^4$ walker positions of the converged chains we compute the GCMF (equation~\ref{eq:evsch}) and the initial GCMF (equation~\ref{CIMF}) from $\Delta$ and $\Mc$ and at each mass we determine the {$[5\%, 50\%, 95\%]$} percentiles of the initial and present-day GCMF. The full-blue and dashed-green lines show the 50\% (i.e. median) values for the GCMF and initial GCMF, respectively, while the shaded regions indicate the {90\% credible} intervals. The fit results are similar to what was found by \cite{2007ApJS..171..101J} for the 1996 Harris catalogue: $\log\Delta = 5.4\pm0.1$ and $\Mc=5.9\pm0.1$. Note that Jord\'{a}n et al. did not consider stellar evolution mass loss, so their initial GCMF was truncated at $\Mc$, while ours is truncated at $2\Mc$. 

We also compute $K$ with equation~(\ref{fml}) for these $10^4$ walker positions and find {$K=32.5^{+86.9}_{-17.7}$ (90\% credible)}. The spread in $K$ provides an estimate of the uncertainty in $K$, given the  156  Milky Way GC masses. The merger rate will not increase by the same factor of $K$. This is firstly  because half of the value of $K$ is due to stellar mass loss. {The decrease in GC population mass by evaporation is {$K_\Delta = K/2 = 16.3^{+43.5}_{-8.87}$}. Using the approximation for the number of mergers in the observable redshift range from \cite{2020MNRAS.492.2936A0}, $\nmerge \propto M_0^{1.6}\rhn^{-0.67}$, we can estimate the fractional increase in the merger rate ($\fmerge$) as a function of $K_\Delta$. The relation between $\fmerge$ and $K_\Delta$ depends on the adopted mass-radius relation.
 If we parameterise this as $\rhn\propto M_0^\mu$, then we find that for  $\mu=0$ (i.e. a constant initial radius) that then {$\fmerge\simeq3.5$}; for  $\mu =1/3$ (i.e. a constant initial half-mass density) we find {$\fmerge\simeq5$} and for $\mu=0.6$ (i.e. a Faber-Jackson-like relation, \cite{2005ApJ...627..203H,2010MNRAS.408L..16G}) we find {$\fmerge\simeq8$}. The reason that $\fmerge$ increases with $\mu$, is because for large $\mu$, the low-mass clusters are denser and  produce more BHB mergers. 
Because we do not know the initial mass-radius relation, the value of $\fmerge$ is thus in the range $2.4-17.6${, corresponding the range in $K_\Delta$ mentioned above: $10.1-31.2$}. }

Previous studies adopted a constant $K=2.6$ to account for evaporation \cite{2018ApJ...866L...5R, 2018PhRvD..98l3005R} and then assumed that $\fmerge=K$. This value for $\fmerge$ is on the lower boundary of our estimated range {of $K_\Delta$ for} clusters with a constant radius, {corresponding to a factor of $\sim2$ below the  lower boundary of the distribution of $K$}. For our upper boundary of {$K_\Delta$ for} $\mu=0.6$ this value of $\fmerge$ is a factor of $6.8(13.4)$ {lower than our $K_\Delta(K)$}.

\subsection{The GC formation rate}

Next, we  compute
the cluster formation rate $\rhogcdot\equiv \rhogcdot(\tau)$, {where $\tau$ is lookback time}, for a set of model assumptions.
We do this in terms of  a normalised GC formation rate $\rgc\equiv\rgc(\tau)$, such that $\int_\infty^0 \rgc\dr \tau = 1$. {In the next section we will derive $R$ from a model for GC formation across cosmic time.}
For a given present-day $\rhogc$ (equation~\ref{eq:rhogc}), and no mass loss, the GC formation is then found from 
 $\rhogcdot=\rhogc \rgc$.  We now show how $\rhogcdot$ can be easily derived for a population of GCs with a given present-day mass function  that have lost  mass by stellar evolution and/or escape.
Thus, after we determine $\phicln$, we can find $\rhogcdot$ from imposing 
\begin{equation}
  \rhogcdot =     K \rhogc R .
\label{eq:rhogcdot}
\end{equation}

The cluster mass formed  per unit volume integrated over all times  is 
\begin{align}
  \rhogcn &= \int_\infty^0 \rhogcdot\dr \tau= K\rhogc,\\
   &=  2.4^{+2.3}_{-1.2}\times10^{16}~\msun\,\gpc^3\ .
   \label{eq:rhogcn}
\end{align}
The large error bars are because of the uncertainty in $K$ {and $\langle\eta\rangle$}, and imply that $\rhogcn$ is uncertain by a factor of 2. In the next section we include this uncertainty in the predictions for the merger rate.

\section{Methodology} \label{clev} 
The evolution of the BHBs in our cluster models is computed using the
fast code {\modelbbr}. While the details of this method are described in \citet{2020MNRAS.492.2936A0},
here we give a brief summary of the model philosophy, including the full set of differential equations that are used to compute the secular evolution of the cluster models and the merging BHBs they produce.

\subsection{\model}
We assume that the cluster consists of two types of members: BHs and  {  all the other members} (i.e. other stellar remnants and stars). Each contribute a total mass of $\Mbh$ and $\Mst$, respectively, such that the total cluster mass is $\Mcl=\Mst+\Mbh$.
We assume that
after several relaxation time-scales the cluster reaches a state of balanced evolution \citep{1975IAUS...69..133H,2013MNRAS.432.2779B},
so that  the heat generated in the core by the BHBs and the evolution of the  cluster global  properties are related as \citep{1961AnAp...24..369H,2011MNRAS.413.2509G,2013MNRAS.432.2779B}:
\begin{equation}\label{edot}
\dot{E} = \zeta\frac{|E|}{\trh}, 
\end{equation}
where $E\simeq-0.2G\Mcl^2/\rh$ is the total energy of the cluster, with $\Mcl$ the total cluster mass and $\rh$ the half-mass radius. The constant $\zeta\simeq0.1$ \citep{2011MNRAS.413.2509G},
 and $\trh$ is the average relaxation time-scale within $\rh$ which is given by \citep[e.g.,][]{1971ApJ...164..399S}
 \begin{equation}\label{rel}
\trh = 0.138\sqrt{\frac{\Mcl\rh^3}{G}}\frac{1}{\mmean\psi\ln\kappa}.
\end{equation}
Here $\mmean$ is the  mean mass of the stars and remnants {(initially $\mmean=0.638\,M_\odot$)}, and $\ln\kappa$ is the Coulomb logarithm, which varies slowly with $N$, but we fix it to $\ln\kappa=10$. The quantity $\psi$ depends on the mass spectrum within $\rh$, for which we adopt the following form:
\begin{equation}
\delta=1+a_1{f_{\rm BH}} \ ,
\end{equation}
where $f_{\rm BH}=M_{\rm BH}/\Mcl$ is the fraction of 
mass in BHs to the total cluster mass {and $a_1$ is a constant that was determined from a comparison to $N$-body models (see below)}.
We define the start of the balanced evolution as
\begin{equation}
\tcc = \ntrh\trhn.
\end{equation}

Under the above assumptions, the set of coupled ordinary differential equations given below are integrated forward in time to obtain solutions for $\Mbh(t)$, $\Mcl(t)$ and $\rh(t)$. 

The mass loss rate of BHs is coupled to the energy generation rate, which itself is coupled to the total $E$ and $\trh$ of the cluster (equation~\ref{edot}), such that \citep[][]{2013MNRAS.432.2779B}
\begin{equation}\label{eq:mbhej}
\Mbhdot=
\begin{cases}
0,&t<\tcc{\rm ~or~}\Mbh=0,\\
\displaystyle -\beta  {\Mcl\over \trh}\ , & t\ge\tcc{\rm~and~}\Mbh>0 .
\end{cases}
\end{equation}
The cluster mass loss 
due to stellar evolution  is 
\begin{equation}
\Mstdotsev=
\begin{cases}
0, &t<\tsev,\\
\displaystyle -\nu {\Mst\over t}, &t\ge \tsev,
\end{cases}
\label{eq:Mstdotsev}
\end{equation}
with $\tsev\simeq 2\,\rm Myr$.
We  include here an additional (mass independent) term which was not present in \citet{2020MNRAS.492.2936A0}, and that accounts
for cluster evaporation 
\begin{equation}\label{mlt}
\dot{M}_{\star,\rm ev}=-{\Delta\over \langle t \rangle}\ ,
\end{equation}
with $\langle t \rangle\simeq10\,$Gyr  the averaged cluster formation time. 
The total mass loss rate of the cluster is then
\begin{equation}
\Mcldot=\Mstdotsev+\dot{M}_{\star,\rm ev}+\Mbhdot \ .
\end{equation}

{In balanced evolution,} the expansion rate of the cluster radius as the result of relaxation is
\begin{equation}\label{rer}
\rhdotrlx=\zeta  {\rh\over \trh} + 2\frac{\Mcldot}{\Mcl}\rh\ .
\end{equation}
{Before balanced evolution}, the cluster radius 
expands adiabatically as the result of stellar mass loss at a rate
 \begin{equation}
\rhdotsev=-{\Mstdotsev\over\Mcl}\rh.
\label{eq:rhdotsev}
\end{equation}
The final expression for the half-mass radius evolution is 
 \begin{equation}
\rhdot=
\begin{cases}
\rhdotsev, & t<\tcc,\\
 \rhdotsev + \rhdotrlx , & t\ge\tcc.
\end{cases}
\end{equation}

The remaining parameters were obtained in \citet{2020MNRAS.492.2936A0} by fitting the results of  $N$-body simulations:
 {  $\ntrh = 3.21$, $\beta = 2.80\times10^{-3}$, $\nu = 8.23\times10^{-2}$ and $a_1 = 1.47\times10^{2}$}. 

 \subsection{\modelc}

 The initial contraction of the cluster core due to two-body relaxation leads
 to high central densities of BHs which favor the formation of binaries  through three-body processes. The energy produced by the BHBs reverts the contraction process of the core and powers the subsequent expansion of the cluster as described by equation~(\ref{rer}). We can then
 relate the BHB hardening rate to the rate of energy generation
\begin{equation}\label{ebin}
\dot{E}_{\rm bin}=-\dot{E} \ ,
\end{equation}
where  $\dot{E}_{\rm bin}$ is  the hardening rate of all core binaries.
Equation~(\ref{ebin}) allows us to couple in a simple way the evolution
of the BHBs to the evolution of the cluster model.
It is  important to stress that the
hardening rate equation~(\ref{ebin})
depends neither on the number of binaries present in the cluster core nor on the exact mechanism leading to their formation.

In order to compute a merger rate and the binary properties from equation\ (\ref{ebin}), we need to further specify
the dynamical processes that lead to the hardening and merger of the binaries.
We are interested in mergers {that} occur
through (strong) binary-single dynamical encounters in the cluster core \citep[e.g.,][]{2004ApJ...616..221G,2017arXiv171107452S}.
Thus we consider: (i) mergers that occur in between binary-single encounters while
the binary is still bound to its parent cluster ({in-cluster inspirals}); (ii) mergers
that occur during a binary-single (resonant) encounter as two BHs are driven to a
short separation such that gravitational wave (GW) radiation will lead to their
merger ({GW captures}); and (iii)  mergers that occur after the binary is ejected from its
parent cluster. We use \modelc\ to determine the rate and masses of the BH
binary mergers produced by these three dynamical channels.

{ In balanced evolution,  after a binary is ejected or merges  a new binary must quickly form to meet the energy demand from the cluster. Under such conditions, the binary formation rate  nearly equals the binary ejection rate}
and it is given therefore
by the  BH mass ejection rate equation~(\ref{eq:mbhej}) divided by
the total mass ejected by each binary
\begin{align}\label{rateF}
\Gamma_{\rm bin}&\simeq - \frac{\dot{M}_{\rm BH}}{ m_{\rm ej}},\nonumber\\
\end{align}
where $m_{\rm ej}$ was computed using equation (38) in \cite{2020MNRAS.492.2936A0}, and it is
approximately a fixed number $m_{\rm ej}\simeq 6m$.
The   number of BHBs  that merge before a time $t$ from the formation of the cluster is \citep{2020MNRAS.492.2936A0}
\begin{align}
\mathcal{N}(<t)&=\int_{0}^{t}\Gamma_{\rm bin} 
\left[{{P}}_{\rm in}+{{P}}_{\rm ej}(t-t') \right] \dr t',
\label{eqN}
\end{align}
where $P_{\rm ej}(t-t_{\rm ej})$  is the probability that a binary ejected dynamically from the cluster at
a time $t_{\rm ej}$
will merge due to GW emission within a time $t$ from the formation of the cluster, and
${{P}}_{\rm  in}$ is  the probability that a binary merges inside the cluster.
{ Specifically, ${{P}}_{\rm  in}$ is the sum of the probability that 
that a binary merges through an in-cluster inspiral and the probability of a GW capture.
As described in \cite{2020MNRAS.492.2936A0}, the probability that a binary merges in between two binary-single interactions is given by
integrating the differential merger probability per binary-single encounter over the total number of
binary-single interactions experienced by the binary. Similarly, the probability that a binary merges through a GW capture
is obtained by dividing each binary-single encounter into 20 intermediate resonant
states as in \cite{2017arXiv171107452S}, and by integrating the differential merger probability per resonant encounter over all encounters experienced by a binary.} The merger probabilities are computed by assuming that the   eccentricity of the BHBs follows that  
of a so-called thermal distribution $N(e)\propto e$ \citep[e.g.,][]{Heggie1975},
and their full expressions can be found in \citet{2020MNRAS.492.2936A0}.
Moreover, when evaluating $P_{\rm ej}$ and $P_{\rm in}$ we  set $E_{\rm bin}=-Gm_1m_2/2a$, with $a$ the binary semi-major axis and
$m_1$ and $m_2$ the mass of the BH components, i.e., we have assumed that only one BHB is responsible for all the heating at any given time.
However,  because the dependence is weak, e.g. $P_{\rm in} \propto \dot{E}_{\rm bin}^{-2/7}$ \citep{2020MNRAS.492.2936A0}, and the number of hard
binaries is expected to be of order unity in the type of clusters we consider \cite{2008gady.book.....B},  this simplification is reasonable.

\subsection{Black hole mass function and natal kicks}
In order to calculate the merger rate  through equation~(\ref{eqN})
 we need a physically motivated  model for the BH mass function and
its time evolution.

We sample 100 stellar progenitor masses from the initial mass-function
$\phi_\star(m_\star)\propto m_\star^{-2.3}$ \citep{2001MNRAS.322..231K} between $m_{\star,\rm lo}$  and $100M_\odot$,  with
$m_{\star,\rm lo}\simeq20\,M_\odot$ the stellar mass above which a BH forms.
The resulting masses of the BHs are then obtained using the fast stellar evolution code
SSE \citep{2000MNRAS.315..543H} which we modified to include up to date prescriptions
for stellar wind driven mass loss \citep{Vink2001}, compact-object formation and supernova kicks \citep{2010ApJ...714.1217B,Fryer2012},
and we also include prescriptions
to account for pulsational-pair instabilities and pair-instability supernovae \citep{2016A&A...594A..97B}.
The initial mass fraction in BHs is set equal to  the total mass in BHs divided by the total mass in stars between $0.1$ and $m_{\star,\rm lo}$
for a  \citet{2001MNRAS.322..231K} initial mass function, and ranges from  $f_{\rm bh}\simeq 0.04$ to $0.07$ depending on the metallicity and the adopted prescription 
for compact-object formation.
These mass fractions  {first  increase as the result of stellar evolution mass loss and then they reduce }  due to the ejection of BHs caused by natal kicks and dynamical ejections.
The BH natal kicks are computed using a standard fallback  model in which  the BHs receive a kick drawn from a Maxwellian distribution with dispersion
$\sigma=265\,\kms$ \citep{Hobbs2005}, lowered by the fraction of the ejected supernova mass that falls back into the compact-object.
The fallback fraction and remnant masses are determined according to
  the chosen remnant-mass prescription. We adopt here
  the rapid  supernova prescription described in Section 4 of \citet{Fryer2012},
in which the explosion is assumed to occur within the first 250ms after bounce.
  But later in Section~\ref{Mpar} we also explore other choices for the compact-object
  formation recipe.

The cluster  dynamically processes  its BH population such
that the mass of the merging BHBs
progressively decreases with time because the
most massive BHs are the first to reach the cluster core, form hard binaries
and merge \citep[e.g.,][]{Rodriguez2016a}. Assuming that the merger products
of BHB mergers are ejected, simulations of dense star clusters also show that the merging BHBs have a distribution
of mass-ratio that is strongly peaked around one \citep[e.g.,][]{2017MNRAS.469.4665P}.
Thus, we assume that the BHs taking part in the dynamical interactions have the same mass,
and that at any given time this mass is equal to that of the largest
BH in the cluster, i.e., $m_1=m_2=m_3=m_{\rm max}$. 
The value of $\mmax$ at a given time, can be easily linked to
the time evolution of the total mass in BHs given by \model .
For a generic BH mass function $\phi_\bullet$, we use the fact that
\begin{align}
  \int_{\mlo}^{\mmax}\phi_\bullet m\dr m = \Mbh\ ,
  \label{eq:Mbhint}
\end{align}
where the integral in the left hand side  is simply the
cumulative  distribution of BH masses computed with SSE.
We  then invert numerically this relation to find $\mmax(\Mbh)$.

\subsection{Cluster formation and initial properties}
We sample the initial cluster  masses from the
Schechter mass function equation (\ref{CIMF}), i.e., we assume that both evaporation and stellar mass loss 
are  important. The values [$\Delta$, $\Mc$] {needed to compute $\phicln$ and $K$}
are sampled  from their posterior distributions derived in Section~\ref{GCMF}.

An important property of
 balanced evolution is that the value of a cluster half-mass radius
today is largely independent of its initial value \cite{1961AnAp...24..369H,2010MNRAS.408L..16G}.
It is therefore not possible to infer the initial density of GCs from their properties today.
We therefore consider three choices for the initial stellar mass density within the half-mass radius, $\rho_0=3M_0/(8\pi{r}_{\rm h,0}^3)$,  which we set
equal to {$\rho_0=10^4\,\mspc$ for our canonical Mod1, and increase(decrease) by a factor 10 in Mod2(Mod3) to explore the effect of initial cluster density.}

For the cluster metallicity,  we fit a quadratic polynomial to the observed age-metallicity relation for
the Milky Way GCs \citep{VandenBerg2013}, to obtain the mean metallicity
\begin{equation} \label{mmet}
\log(Z_{\rm mean}/Z_\odot)\simeq0.42+0.046\left({t\over {\rm Gyr}}\right)-0.017\left({t\over {\rm Gyr}}\right)^2 \ .
\end{equation}
Given the cluster age, $t$, we then assume a log-normal distribution of metallicity around the mean,
with standard deviation $\sigma= 0.4$\,dex. This
takes into account the large spread found in the observed age-metallicity relation. In order to determine
the effect of metallicity on our results  we will later consider additional models where the cluster metallicity is set to a
fixed value.

We  obtain the distribution of cluster formation times
from the semi-analytical galaxy formation model of \citet{2019MNRAS.482.4528E}.
The same model  has also been used in  recent work 
\citep[e.g.,][]{2018ApJ...866L...5R,2019PhRvD.100d3009S}, allowing a direct comparison of our results to  {} literature. \citet{2019MNRAS.482.4528E}
describe
the process of GC formation as resulting from  star formation activity in the high-density disks of gas-rich
galaxies. Motivated by the results of simulations of molecular cloud collapse, they assume that massive bound clusters form
preferentially when the gas surface density exceeds a certain threshold. 
Applying this Ansatz to a semi-analytic gas
model built on dark matter merger trees, they
make  predictions
for the cosmological formation rate of GCs. { The resulting cluster formation history peaks at a redshift of $\sim4$, which is earlier than the peak in the cosmic star formation history (redshift $\sim2$, \cite{Madau2014}). } We sample the formation redshift of our cluster models from the total cosmic cluster formation rate given by 
the fiducial model of \citet{2019MNRAS.482.4528E}
  and integrated over all halo masses. This corresponds to  the formation rate per comoving volume of their Figure~8 with
their parameters $\beta_\Gamma=1$ and $\beta_\eta=1/3$, where $\beta_\Gamma$ sets the dependence of the cluster formation efficiency on surface density, and $\beta_\eta$ the dependence
of the star formation rate on the halo virial mass.
We then normalize the GC formation to unit total number to obtain $R$ (equation~\ref{eq:rhogcdot}).
Thus, we only sample the cluster formation redshift from the \citet{2019MNRAS.482.4528E} model, while the total
 cluster {formation rate is given by our equation~(\ref{eq:rhogcdot}).}
{ For this model, approximately 25\% of the cosmic star formation \cite{Madau2014} is in star clusters at redshifts $\gtrsim 4$ (i.e. before the peak) for  $K=32.5$, implying that $K=130$ is an upper limit to ensure that the cluster formation rate is below the star formation rate. This limit corresponds to $2.5\sigma$ in our $K$ distribution and hence it is unlikely to occur. }
Later, in order to determine the importance of our assumption about the cluster formation history, we will  consider another class of models with different values for $\beta_\Gamma$ and $\beta_\eta$.

Given the initial cluster mass, radius, metallicity, and formation time, the BHB merger rate, $\dot{\mathcal{N}}$,
is  obtained from  \modelbbr .

\section{binary black hole  merger rate}\label{MR}
The merger rate density of BHBs at a lookback time $\tau$ is 
\begin{eqnarray}
  \mathcal{R}(\tau)&=&\int\int\int { \phicln }(M_0) \rgc(\tau') s(Z) \nonumber \\ &&\dot{\mathcal{N}}(\tau'-\tau; M_0, \rhn, Z)
\dr \tau' \dr M_0 \dr Z
\label{eq:Rtau}
\end {eqnarray}
where $\dot{\mathcal{N}}(t; M_0, \rhn, Z)$ is the BHB merger rate corresponding to a cluster
with an initial mass $M_0$, half-mass radius $\rhn$ and metallicity $Z$
at a time $t$ after its formation; $\rgc(\tau)$ is the normalized cluster formation rate and  $s(Z)$ is the normalized  formation rate of clusters with
a metallicity $Z$ at a time $\tau$, $\int s(Z;\tau)\dr Z=1$, which  can be
calculated  given a  model for the time evolution of metallicity, e.g., equation~(\ref{mmet}).

\begin{figure}
    \centering
    \includegraphics[width=3.1in,angle=0.]{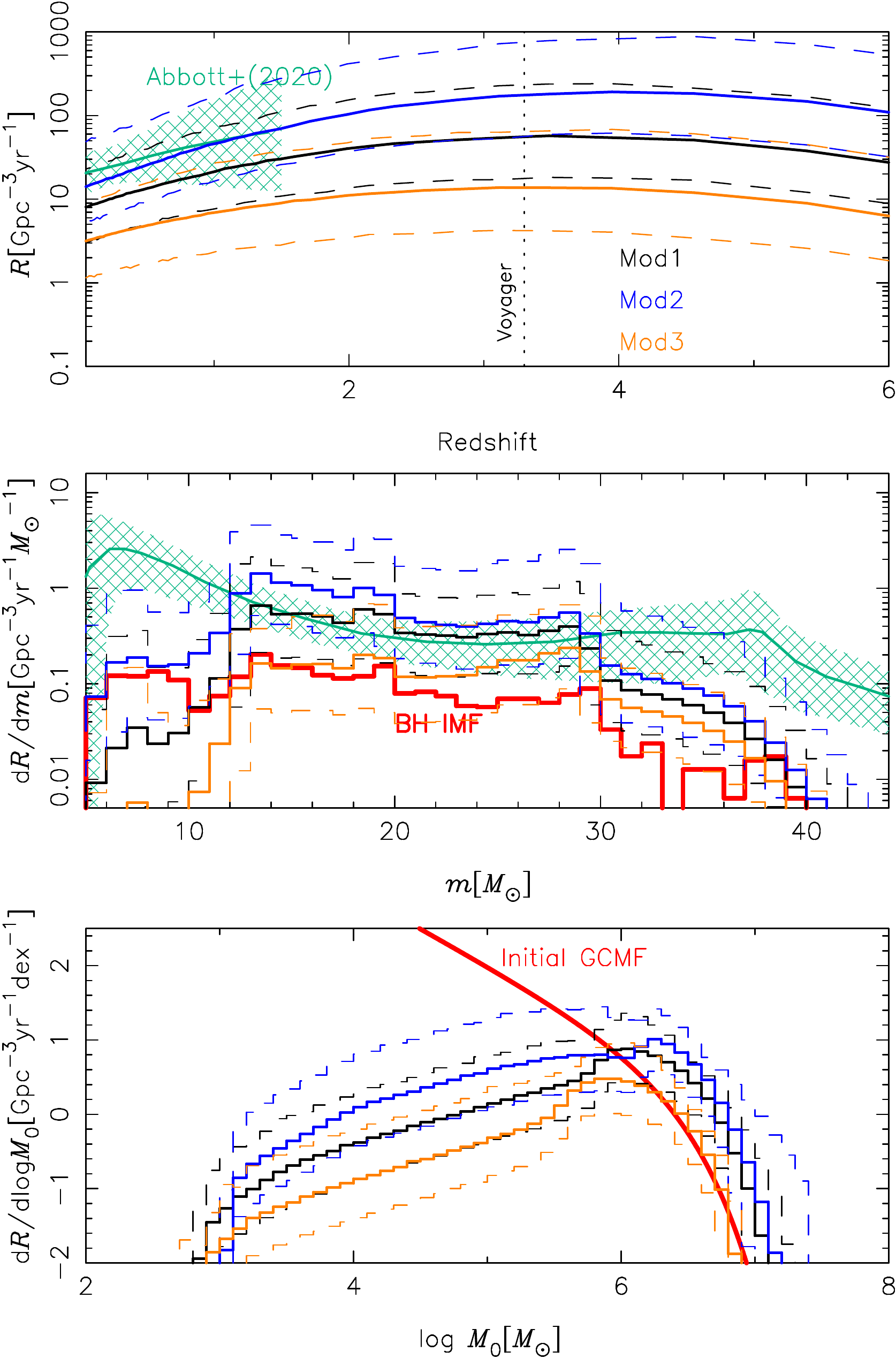}
    \caption{Median of the merger rate distribution (solid lines) for  different  initial cluster half-mass densities corresponding to Mod1,  Mod2 and Mod3 in Table~\ref{T1}. The dashed lines contain 
$90\%$ of all model realizations (between the $5$ and $95$ percentiles). The upper panel gives the merger rate as a function of redshift. The  middle and lower panels show the differential local merger rate as a function of the the primary BH mass and initial cluster mass, respectively. We compare our results
      to the median (solid), and the $90\%$ credible intervals (hatched regions) inferred from the GWTC-2 catalogue in
      \cite{Abbott:2020gyp}.
      In the middle panel, we have used their {`Power Law \& Peak'} model and
      the thick-red line  gives the BH
      initial mass function (in arbitrary units). In the lower
      panel we show the initial GMF (in arbitrary units) four our best fit value $\log(2M_c/M_\odot)=6.26$.
 Vertical dotted line corresponds to the LIGO Voyager upgrade horizon for $(10+10)~M_\odot$ BHBs \citep{2017CQGra..34d4001A}.
      The observable horizons of the Einstein telescope \citep{Punturo_2010} and the Cosmic Explorer \citep{2017CQGra..34d4001A} extend to the
very early Universe and are both to the right of the $x$-axis range in the figure.
    } 
    \label{fig2}
\end{figure}

\begin{figure}
    \centering
    \includegraphics[width=3.4in,angle=0.]{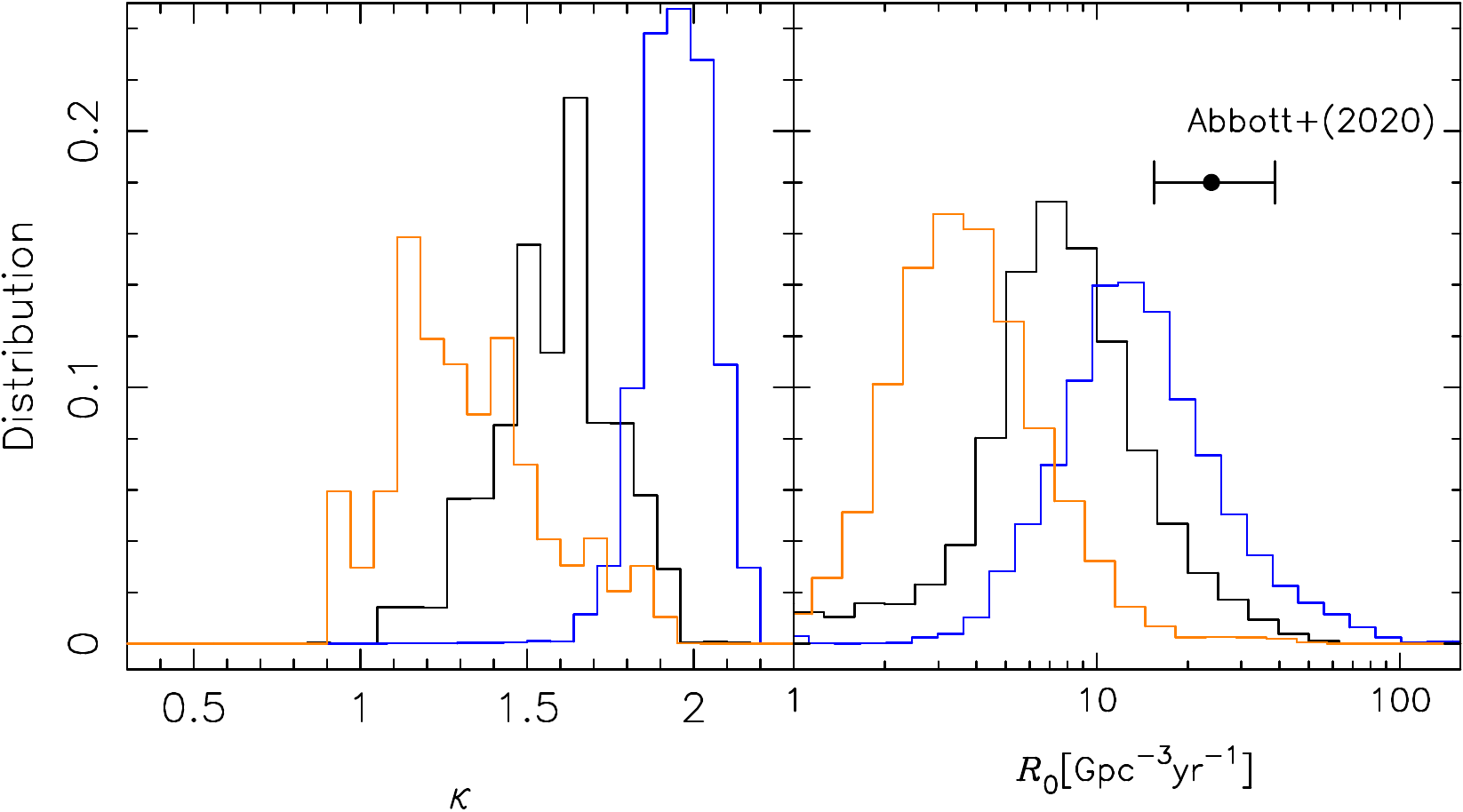}
    \caption{Distribution of the rate parameter $\kappa$, and the local
      merger rate, $\mathcal{R}_0$, for each of the three models of Fig.~\ref{fig2}.      
     Colors are as in Fig.\ref{fig2}.
    } 
    \label{fig3}
\end{figure}

\begin{table*}
\centering
\begin{tabular}{l| cccccc|ccc}
\hline
Model & Density  & Z & Cluster formation &  BH masses & Natal & $\dot{M}_{\star,\rm ev}$ &$\mathcal{R}_0$  & $\kappa$ &$\mathcal{R}_{0, \rm in-cluster}$  \\
 & $\mspc$&  &  &   & kicks &  & [{$\rm Gpc^{-3}yr^{-1}$}] &  & [{$\rm Gpc^{-3}yr^{-1}$}] \\
\hline\hline

Mod1 & $10^4$  & Eq. (\ref{mmet}) & $\beta_\Gamma=1;~\beta_\eta=1/3$   & Rapid & fallback & Eq. ~(\ref{mlt})& $7.2^{+12.7}_{-4.5}$ & $1.61^{+0.26}_{-0.37}$
& $1.8^{+1.6}_{-1.0}$\\
Mod2 & $10^5$ & -& - &      -& - & - & $12.2^{+32.9}_{-7.9}$ & $1.92^{+0.16}_{-0.17}$ & $2.0^{+1.7}_{-1.1}$\\
Mod3 & $10^3$ &  -& - &- & -  & - &$3.0^{+5.6}_{-1.9}$  &$1.23^{+0.53}_{-0.35}$ & $1.1^{+1.2}_{-0.6}$\\
\hline
Mod4 & $10^4$ & 0.01$\rm Z_\odot$ & - & - & - &  - &$6.8^{+13.6}_{-3.9}$ & $1.60^{+0.22}_{-0.33}$ & $1.8^{+1.8}_{-1.1}$\\
Mod5 & - & 0.1$\rm Z_\odot$  & - & - & - & - &$7.5^{+16.0}_{-5.3}$ & $1.57^{+0.27}_{-0.38}$ & $2.0^{+2.0}_{-1.2}$ \\
Mod6 & - & $\rm Z_\odot$  & - & - & -  & - &$6.9^{+15.4}_{-3.2}$  &$1.51^{+0.30}_{-0.26}$ &  $2.7^{+2.5}_{-1.7}$\\
Mod7 & - & Eq. (\ref{mmet})  & All form at z=3 & - & - & - &$7.5^{+15.5}_{-4.5}$ & $1.72^{+0.26}_{-0.34}$ & $1.8^{+1.8}_{-1.1}$\\
Mod8 & - & - & $\beta_\Gamma=0;~\beta_\eta=1/3$&- & - & - & $7.6^{+16.3}_{-5.1}$ & $1.53^{+0.25}_{-0.38}$ & $2.1^{+2.5}_{-1.4}$\\
Mod9 & - & - & $\beta_\Gamma=1;~\beta_\eta=1/6$&- & - & - & $6.6^{+13.6}_{-4.6}$ & $1.54^{+0.29}_{-0.29}$ & $1.8^{+1.8}_{-1.0}$\\
Mod10 & - & - & $\beta_\Gamma=1;~\beta_\eta=1/3$  & \cite{Belczynski2002}  &- & - &$7.1^{+13.6}_{-4.0}$
& $1.68^{+0.22}_{-0.31}$ &$1.5^{+1.3}_{-0.9}$\\
Mod11 & - & -& -  &  \cite{Belczynski2008}         & - & - &$7.8^{+16.1}_{-4.8}$ & $1.64^{+0.30}_{-0.26}$  & $1.7^{+1.5}_{-1.0}$ \\
Mod12 & - & -& - & Delayed   & -  & - & $7.1^{+14.6}_{-4.4}$  &$1.58^{+0.27}_{-0.36}$ & $1.8^{+1.9}_{-1.0}$ \\
Mod13 & - &  - & -  & Rapid& No kicks & - &$10.0^{+21.4}_{-6.9}$ & $1.52^{+0.26}_{-0.31}$ & $2.4^{+2.5}_{-1.4}$\\
Mod14 & - &  - &-   &-& momem.      & - &$8.0^{+14.5}_{-6.1}$  &$1.50^{+0.31}_{-0.34}$ &  $2.2^{+2.2}_{-1.3}$ \\
Mod15 & - & - & - & - & fallback & 0 & $2.2^{+1.4}_{-1.4}$ & $1.22^{+0.13}_{-0.29}$ & $0.9^{+0.2}_{-0.2}$ \\
\hline
Mod16 & $10^5$ &0.01$\rm Z_\odot$ & -& - & - & Eq. ~(\ref{mlt})& $12.2^{+29.6}_{-8.0}$ & $1.87^{+0.14}_{-0.17}$
& $1.9^{+1.9}_{-1.1}$ \\
Mod17 & - & 0.1$\rm Z_\odot$  & - & - & - & - &$12.7^{+32.4}_{-8.1}$ & $1.93^{+0.12}_{-0.20}$ & $2.0^{+1.9}_{-1.2}$ \\
Mod18& - & $\rm Z_\odot$  & - & - & -  & -    &$10.3^{+24.5}_{-6.7}$  &$2.28^{+0.12}_{-0.35}$ &  $2.3^{+2.9}_{-1.5}$\\
Mod19 & - & Eq. (\ref{mmet})  & All form at z=3 & - & - & - &$11.1^{+27.9}_{-7.2}$ & $2.10^{+0.03}_{-0.12}$ & $1.8^{+1.5}_{-1.2}$\\
Mod20 & - & - & $\beta_\Gamma=0;~\beta_\eta=1/3$ &- & - & - & $12.3^{+29.4}_{-8.3}$ & $1.91^{+0.19}_{-0.21}$ & $2.2^{+2.2}_{-1.3}$\\
Mod21 & - & - & $\beta_\Gamma=1;~\beta_\eta=1/6$ &- & - & - & $11.5^{+28.1}_{-7.6}$ & $1.90^{+0.10}_{-0.15}$ & $1.8^{+1.7}_{-1.1}$\\
Mod22 & - & - & $\beta_\Gamma=1;~\beta_\eta=1/3$  & \cite{Belczynski2002}  &- & - &$13.1^{+34.0}_{-8.4}$ & $1.90^{+0.11}_{-0.16}$ &$1.9^{+1.9}_{-1.1}$\\
Mod23 & - & -& -  & \cite{Belczynski2008}         & - & - &$14.5^{+36.7}_{-8.8}$ & $1.88^{+0.13}_{-0.13}$  & $2.0^{+1.8}_{-1.2}$ \\
Mod24 & - & -& - & Delayed   & -  & - & $11.8^{+27.3}_{-6.8}$  &$1.96^{+0.12}_{-0.19}$ & $1.9^{+1.7}_{-1.3}$ \\
Mod25 & - &  - & -  & Rapid& No kicks & - &$18.1^{+36.7}_{-6.1}$ & $1.80^{+0.14}_{-0.17}$ &  $3.1^{+2.9}_{-1.7}$ \\
Mod26 & - &  - &-   &-& momem.      & -  &$14.2^{+32.6}_{-9.2}$ & $1.86^{+0.18}_{-0.20}$ & $2.7^{+2.4}_{-1.7}$\\
Mod27 & - & - & - & - & fallback & 0 & $3.2^{+2.4}_{-2.1}$ & $1.87^{+0.10}_{-0.19}$ & $1.0^{+0.4}_{-0.4}$ \\
\hline
Mod28 & $10^3$ &0.01$\rm Z_\odot$ & - & - & - & Eq. ~(\ref{mlt})& $2.7^{+4.7}_{-1.8}$ & $1.30^{+0.49}_{-0.33}$
& $1.0^{+0.9}_{-0.6}$ \\
Mod29 & - & 0.1$\rm Z_\odot$  & - & - & - & - &$3.4^{+5.8}_{-2.2}$ & $1.16^{+0.52}_{-0.27}$ & $1.1^{+1.2}_{-0.8}$ \\
Mod30& - & $\rm Z_\odot$  & - & - & -  & - &$3.5^{+7.3}_{-2.5}$  &$0.91^{+0.32}_{-0.30}$ &  $1.4^{+2.0}_{-0.8}$\\
Mod31 & - & Eq. (\ref{mmet})  & All form at z=3 & - & - & - &$2.9^{+5.2}_{-1.9}$ & $1.39^{+0.48}_{-0.35}$ & $1.0^{+1.0}_{-0.7}$\\
Mod32 & - & - & $\beta_\Gamma=0;~\beta_\eta=1/3$ &- & - & - & $3.3^{+5.9}_{-2.1}$ & $1.19^{+0.41}_{-0.35}$ & $1.1^{+1.2}_{-0.6}$\\
Mod33 & - & - & $\beta_\Gamma=1;~\beta_\eta=1/6$ &- & - & - & $2.9^{+4.0}_{-1.9}$ & $1.26^{+0.47}_{-0.38}$ & $1.0^{+1.0}_{-0.6}$\\
Mod34 & - & - & $\beta_\Gamma=1;~\beta_\eta=1/3$   & \cite{Belczynski2002}  &- & - &$2.7^{+4.5}_{-1.6}$ & $1.44^{+0.46}_{-0.39}$ &$0.9^{+0.8}_{-0.5}$\\
Mod35 & - & -& -  & \cite{Belczynski2008}         & - & - &$2.9^{+4.9}_{-1.7}$ & $1.39^{+0.47}_{-0.33}$  & $1.0^{+0.9}_{-0.6}$ \\
Mod36 & - & -& - & Delayed   & -  & - & $2.9^{+5.0}_{-1.9}$  &$1.22^{+0.44}_{-0.35}$ & $1.0^{+1.0}_{-0.6}$ \\
Mod37 & - &  - & -  & -& No kicks & - &$3.7^{+6.4}_{-2.5}$ & $1.32^{+0.46}_{-0.31}$ & $1.2^{+1.1}_{-0.7}$\\
Mod38 & - &  - &-   &-& momem.      & - &$2.9^{+3.9}_{-1.8}$  &$1.02^{+0.54}_{-0.34}$ &  $1.1^{+1.0}_{-0.6}$ \\
Mod39 & - & - & - & - & fallback & 0 & $0.8^{+0.6}_{-0.4}$ & $0.62^{+0.21}_{-0.30}$ & $0.46^{+0.02}_{-0.06}$ \\
\hline
\end{tabular}
\caption{Model parameters used in the calculations. Here $\beta_\Gamma$ and $\beta_\eta$
  refer to the  parameters of the cosmological  models in \citet{2019MNRAS.482.4528E} that are used
  to sample the cluster ages in our simulations.
  The rightmost three columns give the local merger rate density of BHBs, the
rate evolution parameter $\kappa$, and the local merger rate of in-cluster mergers (including GW captures).
}\label{T1}
\end{table*}

{ In practice, for each model assumption in Table~\ref{T1} we
sample 100 values over the posterior distributions of the
parameters $\Mc$ and $\Delta$ obtained from the MCMC fit to the observed Milky Way GCMF.
Then, for each  pair [$\Mc$, $\Delta$],
we evolve $N_{\rm cl}=600$ models with masses
sampled over  a grid of  constant logarithmic step size, $\delta \log M/\msun=0.01$, in the range $10^2-10^8\msun$, and use that:
\begin{equation}\label{SUM}
\mathcal{R}(z)\simeq K\rhogc { \sum\limits_{i=1}^{N_{\rm cl}}\dot{\mathcal{N}}(z; M_{0,i}) \phicln(M_{0,i}) M_{0,i} 
\over
\sum\limits_{i=1}^{N_{\rm cl}}\phicln(M_{0,i})M_{\rm 0,i}^2},
\end{equation}
where the formation time of each cluster is randomly sampled from the corresponding $R(\tau)$ distribution; we then  use equation~(\ref{mmet}) to compute the mean metallicity, $Z_{\rm mean}$, that corresponds to that formation time, and thus find the cluster metallicity   by drawing from a  log-normal distribution around 
$Z_{\rm mean}$. 
The half-mass radius of the cluster 
is obtained from the cluster mass given the assumed half-mass density. 
  Note that because 
each  cluster has its own metallicity, each time we generate a new BH population using SSE; 
the fraction of clusters with $Z < 0.1Z_\odot$ is $\simeq 84\%$.
 We also take into account the uncertainty on the  mass density of GCs in the Universe, $\rhogc$.
We assume that the parameter
$\rhogc$ follows a Gaussian distribution with mean $7.3\times 10^{14}M_\odot\ \rm Gpc^{-3}$ and $\sigma=2.6\times 10^{14}M_\odot\ \rm Gpc^{-3}$. We sample 1000 values from this latter distribution and for each of them we use
 equation~(\ref{SUM}) to determine a merger rate estimate for each of the [$\Mc$, $\Delta$] values, 
and thus obtain a {\it distribution} of merger rate density values.

Because our results turn out to be  more sensitive to the cluster initial density than to other parameters,
we first focus on Mod1, Mod2 and Mod3 in Table~\ref{T1}.
In these models we vary $\rho_0$ in a range that is relevant to real globular clusters, while keeping fixed all the other parameters
as given in the table.  
This allows us to bracket a plausible range of
values for the local merger rate density and its redshift evolution.
In Fig.~\ref{fig2} we plot the median of the merger rate distribution and credible intervals
as a function of redshift, and
the primary BH mass distribution  of binaries merging at redshifts $z<1$ as well as the initial mass distribution of
the clusters where these binaries originated.}

Fig.~\ref{fig2} shows that  the difference between our
upper and lower bounds on the comoving BH merger rate density is about a factor $\sim 10$ for any density assumption.
This is due to the fact that
$\mathcal{R}\propto K$
at a very good approximation, and, as we discussed above, 
 tight constraints on $K$ cannot be placed from the
present day Milky Way GCMF. 
Moreover, rates are not too sensitive to the initial cluster density --
two orders of magnitude difference
in the initial density  leads to  a factor of $\sim 5$ variation in the local value of $\mathcal{R}$. {From this we conclude that the  initial density uncertainty is as important as the unknown initial GCMF.  }

For each  initial GCMF, corresponding to new values of [$\Mc$, $\Delta$] and $\langle\eta\rangle$,
we fit the redshift distribution of the merger rate density at $z<2$  using
\begin{equation}\label{rvsz}
  \mathcal{R}(z)=\mathcal{R}_0\left(1+z \right)^{\kappa}\ ,
\end{equation}
to derive  a distribution of values for the parameters $R_0$
and $\kappa$ for each of our three density assumptions.
In this analysis we  neglect the uncertainties associated to each fit because their standard deviations are much smaller
than the variation in the inferred parameters across the different models.
The initial parameters for each of the three densities and the corresponding median values and uncertainties (5 and 95 percentiles)
of $\mathcal{R}_0$ and $\kappa$ are given in Table~\ref{T1} (Mod1, Mod2 and Mod3).
The distributions obtained from this analysis are shown in Fig.~\ref{fig3}.
The local BH merger rate density from GCs varies in the range $R_0\simeq 1$ to
$50 \rm \,Gpc^{-3}yr^{-1}$.
A comparison to  the local merger rate  inferred from the GW detections, $23.9^{+14.9}_{-8.6}\rm Gpc^{-3}yr^{-1}$ { (for their redshift independent results and $19.1^{+16.2}_{-9.0}\,\rm Gpc^{-3}yr^{-1}$ when the merger rate is allowed to evolve with redshift)} \cite{Abbott:2020gyp},
shows that BHBs formed dynamically in GCs are likely to explain a significant fraction of the BHB mergers detected by LIGO-Virgo. Note, however, that
if GCs are formed with high
densities, $\sim10^5\,\mspc$, then our merger rate estimates
are consistent with the LIGO-Virgo
merger rates within uncertainties. { We note that although this high density is preferred to explain the overall rates, the rates in the mass range $13-30\,\msun$ are somewhat better  reproduced  by Mod1 ($10^4\,\mspc$, see Fig.~\ref{fig2}).  }
Combining  the three models together, i.e.,
assuming a universe in which  one third of the clusters form as in Mod1,  one third as in Mod2
and the remaining as in Mod3, we find 
\begin{equation}
\mathcal{R}_0=7.2^{+21.5}_{-5.5}{ \rm Gpc^{-3}yr^{-1}}; ~~ \kappa=1.6^{+0.4}_{-0.6} ,
\end{equation}
where uncertainties refer to the $90\%$ credible intervals.

The middle panel of Fig.~\ref{fig2} shows the primary BH mass distribution
normalized to the   BH merger rate density in the local Universe ($z<1$). Our models MOD1 and MOD2
 reproduce  well the shape as well as the normalization 
of the  BH mass function inferred from the 
GWTC-2 \cite{Abbott:2020gyp}
in the range $13-30\msun$. 
The BH mass distribution appears to be sensitive to  the initial cluster density, in the sense that higher densities lead to
a higher fraction of lower mass BHs among the merging population.
{It is interesting that the higher density models, Mod1 and Mod2, provide a better match to the inferred BH mass function and the rates.} {The additional low-mass BHs in high density GCs} is due to the higher retention fraction of lower mass BHs in  higher density models after a natal kick due to the high escape velocities from these clusters,
 and to the fact that denser clusters process their BH populations faster, thereby `eating' away their BH mass function more. { }

Given the number of complex features that can be seen
in the BH mass distributions we do not attempt here a  parametrization over the full range of BH masses. Moreover,
as we will show later in Section\ \ref{Mpar},
these distributions are  sensitive to the uncertain prescriptions for BH formation, natal kicks and metallicity.
We instead consider the mass range $13\, M_\odot$ to $30\, M_\odot$ where 
a simple power law model, $dR/dm\propto m^\alpha$, does a reasonable job.
In this mass range we
find  $\alpha=0.1^{+0.9}_{-0.5}$ (Mod3), $\alpha=-1.1^{+0.4}_{-0.5}$ (Mod1), and $\alpha=-1.8^{+0.6}_{-0.4}$ (Mod2),
where the reported
values are the median of the distributions  obtained by fitting each of the 100 BH mass
distributions corresponding to  different  [$\Mc$, $\Delta$], and the uncertainties refer to the $5$ and $95$ percentiles.
As before  we  neglect the uncertainties associated to each fit because their standard errors are much smaller
than the variation of $\alpha$ across the different models.
By  combining the three density models together,
 we find
 \begin{equation}
 \alpha =-1.1^{+1.5}_{-1.0} \ .
\end{equation}
Negative values of $\alpha$ are  preferred, though positive values are also
acceptable.
{ The value of the power law index found by us is broadly consistent with the value of $\alpha=-1.58^{+0.82}_{-0.86}$ reported by \cite{Abbott:2020gyp} for their
low-mass ($<40 \msun$)  slope  of the “Broken Power Law”  model}.
For comparison, the BH initial mass function integrated over all metallicities is shown 
in the middle panel of Fig.~\ref{fig2}.
Within the same BH mass range, $m= 13\,M_\odot$ to $30\,\msun$, the best fit power law model to the initial mass function had a spectral index $\alpha\approx-1.8$.
The most striking feature, however, is that 
all mass distributions in Fig.~\ref{fig2} are  strongly depleted at $m\lesssim 15\,M_\odot$.
The fraction of BHs below this mass  decreased by more than one order of magnitude with respect to the BH initial mass function. 
Due to this, all our models underpredict the number of BHBs at $m\sim10 M_\odot$ compared to the mass distribution inferred from the LIGO-Virgo detections.
Moreover, we note  some features that are common to the three models considered here.
All  three models show peaks at $m\simeq 13\,M_\odot$, $20\,M_\odot$ and $30\,M_\odot$.
Above $m=30\,M_\odot$  the BH mass distribution starts to decline rapidly with mass until a break
at $m\simeq 38\,M_\odot$ above which  the decline becomes much steeper.
All models show essentially no BHs with mass above $40\,M_\odot$ or below $5\,M_\odot$.
{  The low merger rate value at $\gtrsim 40\,M_\odot$  is a consequence of the
stellar mass loss prior to the formation of the BHs  because a down-turn above $30\,\msun$ is also seen in the BH IMF and we find it  even in models that do not include any prescription for pair instability \citep{Spera2015a}.
Our pulsational-pair instabilities and pair-instability supernovae prescriptions are taken from \citep{2016A&A...594A..97B}, and for the maximum initial stellar mass  we considered, $100\,M_\odot$, they have little or no effect on the resulting BH masses. 
}
{
Note also that we do not consider hierarchical mergers
 \citep{2016ApJ...831..187A}. 
Their contribution to the merger rate  is sensitive to the distribution of BH natal spins, which is
poorly constrained. 
Assuming that BHs are formed with no spin, \citet{2019PhRvD.100d3027R} finds that $\sim 10\%$ of BHB mergers come from previous mergers;  when the BH dimensionless spin parameter is increased above $0.1$, the contribution drops to only a few per cent or less. { In addition, in the discussion (Section~\ref{ssec:oimbh}) we show that with our adopted mass-radius relation, less 2nd generation mergers are expected compared to \citet{2018PhRvL.120o1101R}}.
Thus, including hierarchical mergers is not expected to significantly change our integrated merger rate estimates. However, the high mass BHs resulting from multiple mergers  can  partly fill up 
 the mass distributions above  
 $m\sim 40\,M_\odot$ where the merger rates from our models are nearly zero.
}

The bottom panel of Fig.~\ref{fig2}  shows the differential contribution of
clusters with different masses to the local merger rate ($z<1$).
This contribution increases with cluster mass until
about  $10^6\,M_\odot$, above which the rate starts to rapidly decrease
because of the exponential truncation of the initial GCMF above $\Mc$.
The contribution of clusters with masses larger than $10^7\,M_\odot$ or smaller
than $10^3M_\odot$ is negligible.
In our models, 
clusters that  have a mass $M_0<4\times 10^5\,M_\odot$ are fully disrupted by the present time. These clusters have been neglected
in some previous work
\citep[e.g.,][]{Rodriguez2015a,Rodriguez2016a,2017MNRAS.469.4665P,2020ApJS..247...48K}, but we find that  they  contribute a
significant fraction of the local merger rate: $\approx 0.33$, $0.48$ and $0.30$
for Mod1, Mod2 and Mod3 respectively.

\subsection{ In-cluster $vs.$ ejected binaries}\label{in}

The merger of a BHB in our models
can occur either after the binary has been ejected dynamically from the cluster,
or within the cluster itself.
In-cluster mergers are  relevant because they can lead to
the formation of BHs with mass above the pair-instability gap at $\approx 50\,M_\odot$ \citep{2016ApJ...831..187A,2018PhRvL.120o1101R,2019MNRAS.486.5008A},
and the observational implications of this have been discussed in a number of previous papers, e.g. \citep{2017ApJ...840L..24F,2017PhRvD..95l4046G,2020arXiv200500023K}.
Among all in-cluster mergers, GW captures are also particularly important because a  fraction of them are expected to have
a finite eccentricity at the moment they first chirp within the 
the LIGO frequency band  above 10 Hz \citep{Antonini2014,Samsing2014,Antonini2015,2017arXiv171107452S,2018PhRvL.120o1101R}.
Thus, they could be identified among other binaries due to their unique eccentric signature.

\begin{figure}
    \centering
    \includegraphics[width=3.2in,angle=0.]{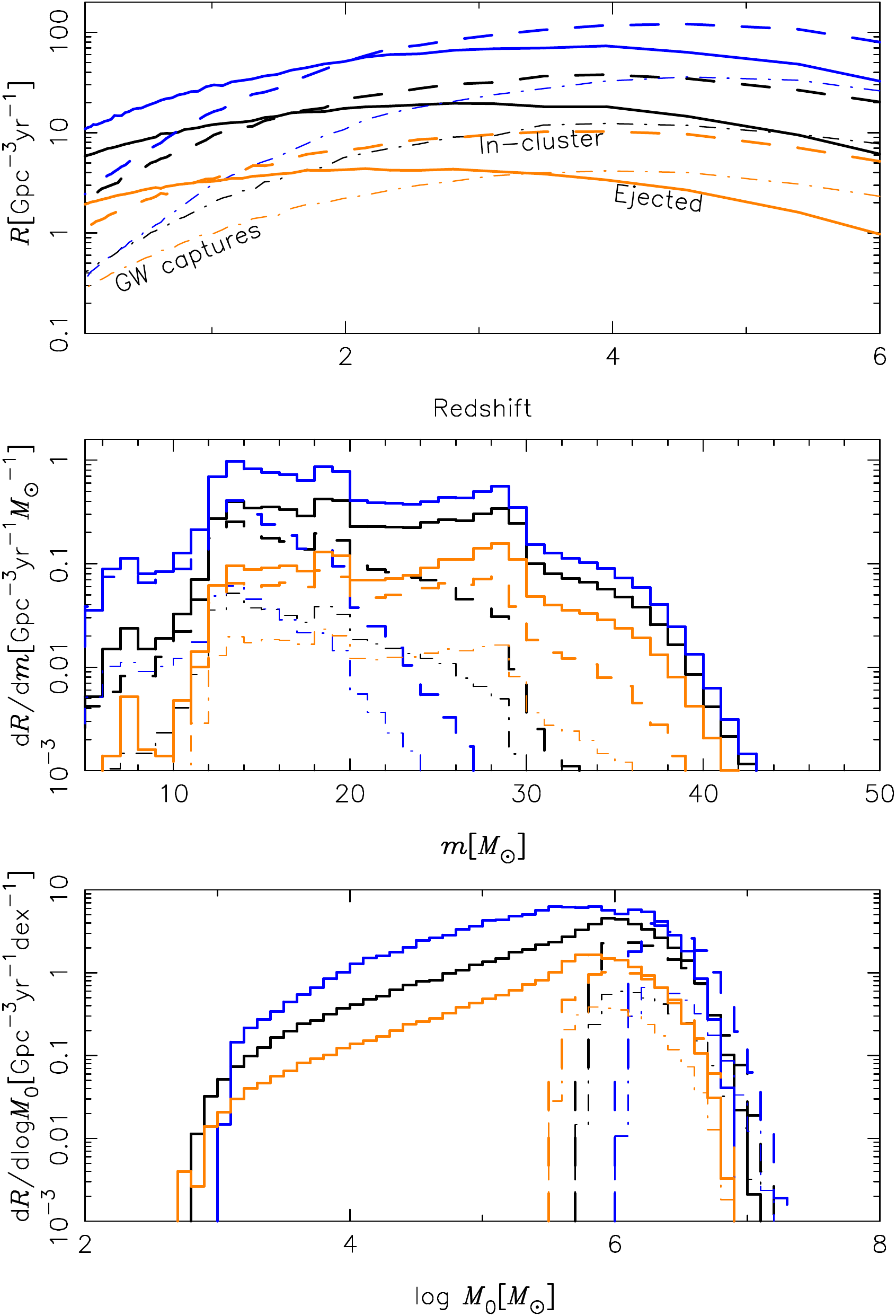}
    \caption{As Fig.\ref{fig2} but we now show separately the median of the  merger rate distribution of all in-cluster mergers (dashed lines), only GW captures (dot-dashed lines), and
      mergers among the ejected binaries (solid lines).     
    } 
    \label{figin}
\end{figure}

In Fig.~\ref{figin} we show separately the  rate evolution of BH mergers
occurring among the ejected binaries, those forming inside the cluster and GW captures, as well as the
mass distributions of mergers at $z<1$ and the mass distribution of their parent clusters.
While  in-cluster mergers dominate the  rate density at early times, $z\gtrsim 2$,
most of the BHB mergers in the local Universe are produced among the ejected  population.
The local rate of in-cluster mergers is  $\simeq 2 \rm\, Gpc^{-3}yr^{-1}$ and that
of GW captures is $\simeq 0.4 \rm \,Gpc^{-3}yr^{-1}$, with little dependence on the initial density assumed.
However, the fractional contribution of in-cluster mergers does depend quite strongly  on the cluster initial conditions,
in the sense that higher  densities lead to lower fractions ---
for $\rho_0=10^3\mspc$, nearly $40\%$ of the local mergers are formed  in-cluster, while for 
the highest initial densities, $\rho_0=10^5\mspc$,   in-cluster mergers only contribute $\simeq 15\%$ of the total. 
The percentage of all in-cluster mergers that occur through a GW capture is
$\approx 20\%$ in the local Universe and increases smoothly with redshift,  reaching $\approx 30\%$
near the peak of cluster formation activity.
Because an order of unity fraction of
gravitational wave captures are expected to have a finite eccentricity ($\gtrsim 0.1$) above 10\,Hz frequency, we conclude 
that eccentric mergers from globular clusters contribute 
$\lesssim 0.4 \rm \,Gpc^{-3}yr^{-1}$ to the  merger rate in the local universe. This low rate is  consistent with the non detection of eccentric binaries in current searches \citep{2019ApJ...883..149A}.

Our models Mod1 and Mod2 show a decrement for the local fraction of in-cluster mergers  over 
previous estimates 
which bracketed this between $30\%$ and $50\%$ of the total rate \citep[e.g.,][]{2018ApJ...866L...5R}.
This difference arises   from the fact that this previous work  only considered clusters with mass 
$\gtrsim 10^5M_\odot$ for which  about half of the overall merger rate
is due to in-cluster binaries.
However, as shown in Fig.~\ref{figin}, lower mass clusters
 contribute significantly to the local rate, although  the mergers they produce 
 only occur among the ejected population. The reason for this is that their BHs
 have all been  ejected by $z=1$.
 Thus, including these systems leads
to an overall reduction of the contribution of  in-cluster mergers and also affects the redshift evolution
of the merger rate density.

\begin{figure}
    \centering
    \includegraphics[width=3.4in,angle=0.]{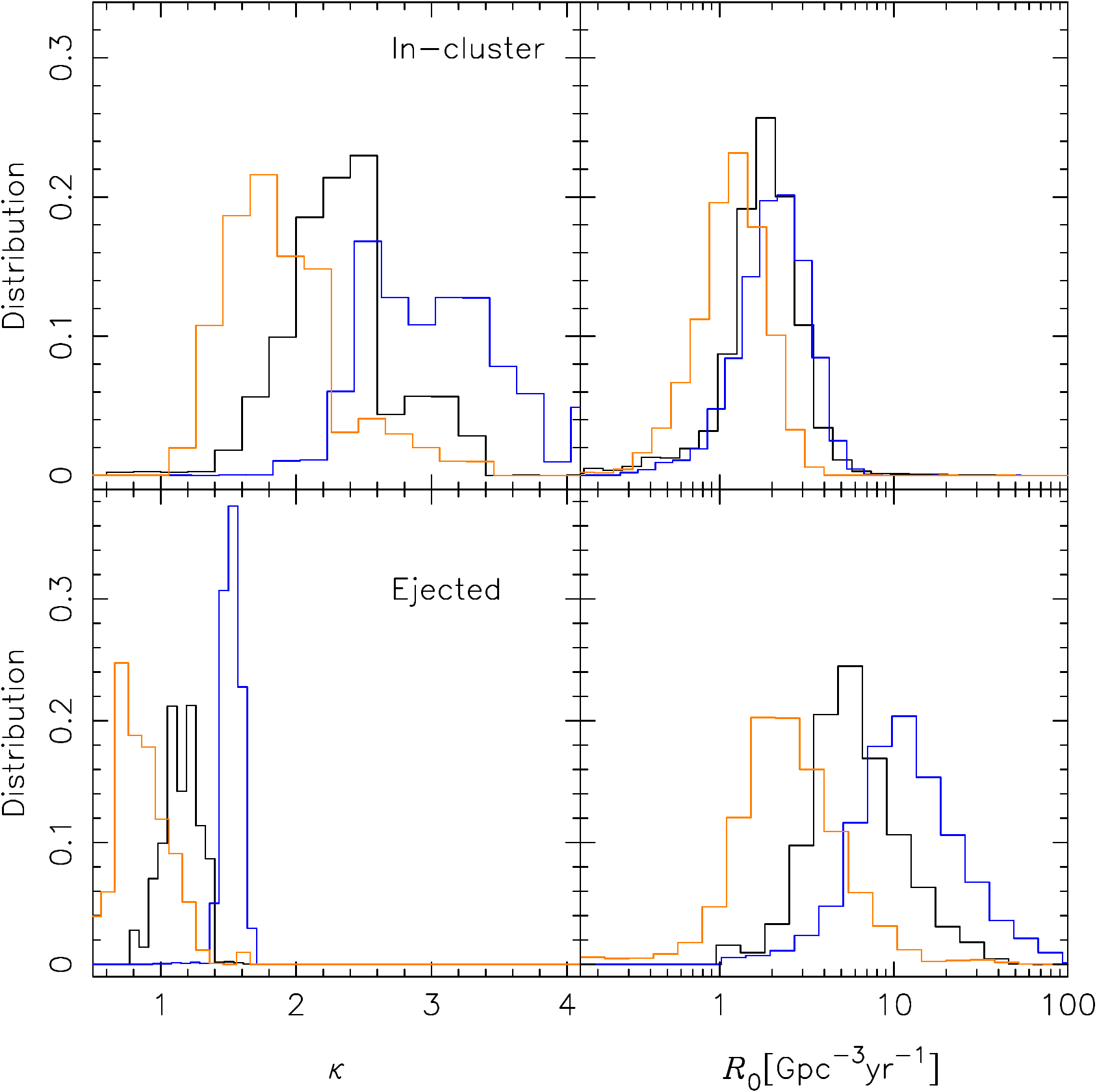}
    \caption{Distribution of the rate parameter $\kappa$, and the local
      merger rate, $\mathcal{R}_0$, for each of the three models of Fig.~\ref{fig2}, and for the in-cluster mergers 
      and ejected binary mergers  separately.
     Colors are as in Fig.\ref{fig2}.
    } 
    \label{figl3}
\end{figure}

Fig.~\ref{figl3} shows the distributions of $R_0$ and $\kappa$ obtained by
fitting equation~(\ref{rvsz}) to the merger rate evolution of in-cluster and ejected binaries
separately. 
The merger rate of in-cluster binaries evolves  steeply with redshift,
 $\kappa \approx 3$, albeit with a large scatter, while for ejected binaries the dependence is much 
weaker, $\kappa\approx 1$.
 If, as before, we assign equal probability to each of our density assumptions
and fit the total merger rate density of in-cluster binaries  using  equation~(\ref{rvsz})
we find 
\begin{equation}
\mathcal{R}_{\rm 0,in}=1.5^{+1.7}_{-0.9}\,{ \rm Gpc^{-3}yr^{-1}}; ~~ \kappa_{\rm in}=2.3^{+1.3}_{-1.0} ,
\end{equation}
while for ejected binaries  
\begin{equation}
\mathcal{R}_{\rm 0,ej}=5.7^{+21.5}_{-4.4}\,{ \rm Gpc^{-3}yr^{-1}}; ~~ \kappa_{\rm ej}=1.2^{+0.4}_{-0.5} .
\end{equation}
We now use a simplified analytical model to gain some physical insights on this result.

From equation~(\ref{eqN}),  the merger rate for
in-cluster binaries is 
\begin{equation}
  {\dr \mathcal{N} \over \dr t}\Big|_{\rm in}=
  \Gamma_{\rm bin} P_{\rm in}(t)~,
\label{deqN2}
\end{equation}
while for ejected binaries we have
\begin{equation}
  {\dr \mathcal{N} \over \dr t}\Big|_{\rm ej}={\dr \over \dr t}\int_{0}^{t}\Gamma_{\rm bin} P_{\rm ej}(t-t') \dr t' ~.
\label{deqN}
\end{equation}
The merger probabilities that enter in the integral equations above can be linked to the evolution of the cluster properties in a simple way under
some simplifying assumptions.
If we neglect cluster mass loss and
that the BHs have a range of masses   -- both have little effect on the merger rate evolution (see Secion~\ref{Mpar}) --
we can write 
 $t_{\rm rh}(t) =t_{\rm rh,0}\left(1+{3\over2}{\zeta t/ t_{\rm rh,0}}\right)$, and $\rho(t) =\rho_0\left(1+{3\over2}{\zeta t/ t_{\rm rh,0}}\right)^{-2}$ \citep{1965AnAp...28...62H, 2019MNRAS.486.5008A}.
 { If $\rh \propto t^{2/3}$ and $M=$constant, then $\rho\propto t^{-2}$, $\vesc\propto t^{-1/3}$ and therefore $\vesc\propto \rho^{1/6}$.  Then from \cite{2020MNRAS.492.2936A0} we know $P_{\rm in}\propto \vesc^{20/7}$, hence $P_{\rm in}\propto \rho^{10/21}$} (neglecting captures) and $P_{\rm ej}\propto(t-t_{\rm ej})^{2/7}\rho(t)^{8/21}$  \cite{2020MNRAS.492.2936A0}.
We can then determine the redshift dependence of the merger rate through equations~(\ref{deqN2}) and (\ref{deqN}).

At times $t\gg t_{\rm rh,0}/{\zeta}$,  for in-cluster binaries  we have 
\begin{equation}
  {\dr  \mathcal{N}\over \dr t }\Big|_{\rm in} \propto (1+z)^{2.9}, 
\end{equation}
where we used that $t(z)\propto (1+z)^{-3/2}$ in order to convert time into redshift.
For ejected binaries we have $ {\dr \mathcal{N} \over \dr t}\Big|_{\rm ej}\propto {t^{-5/7}}$, or
\begin{equation}
  {\dr  \mathcal{N}\over \dr t }\Big|_{\rm ej} \propto (1+z)^{1.1}~.
\end{equation}
Although we have neglected 
some  important ingredients  (e.g., mass loss, BH mass function), the
expected value of $\kappa$ for the two populations is consistent with the ones found above and, as expected, it is much steeper for
in-cluster mergers.
This fits in the view that the rate at which the merging BHBs are produced by a cluster is  controlled
by the relaxation process within the cluster itself,
providing a  physical interpretation to our results. { Moreover it implies
 that most of the merging BHBs at produced by  clusters that are still in the expansion phase.} 

\begin{figure}
    \centering
    \includegraphics[width=1.75in,angle=0.]{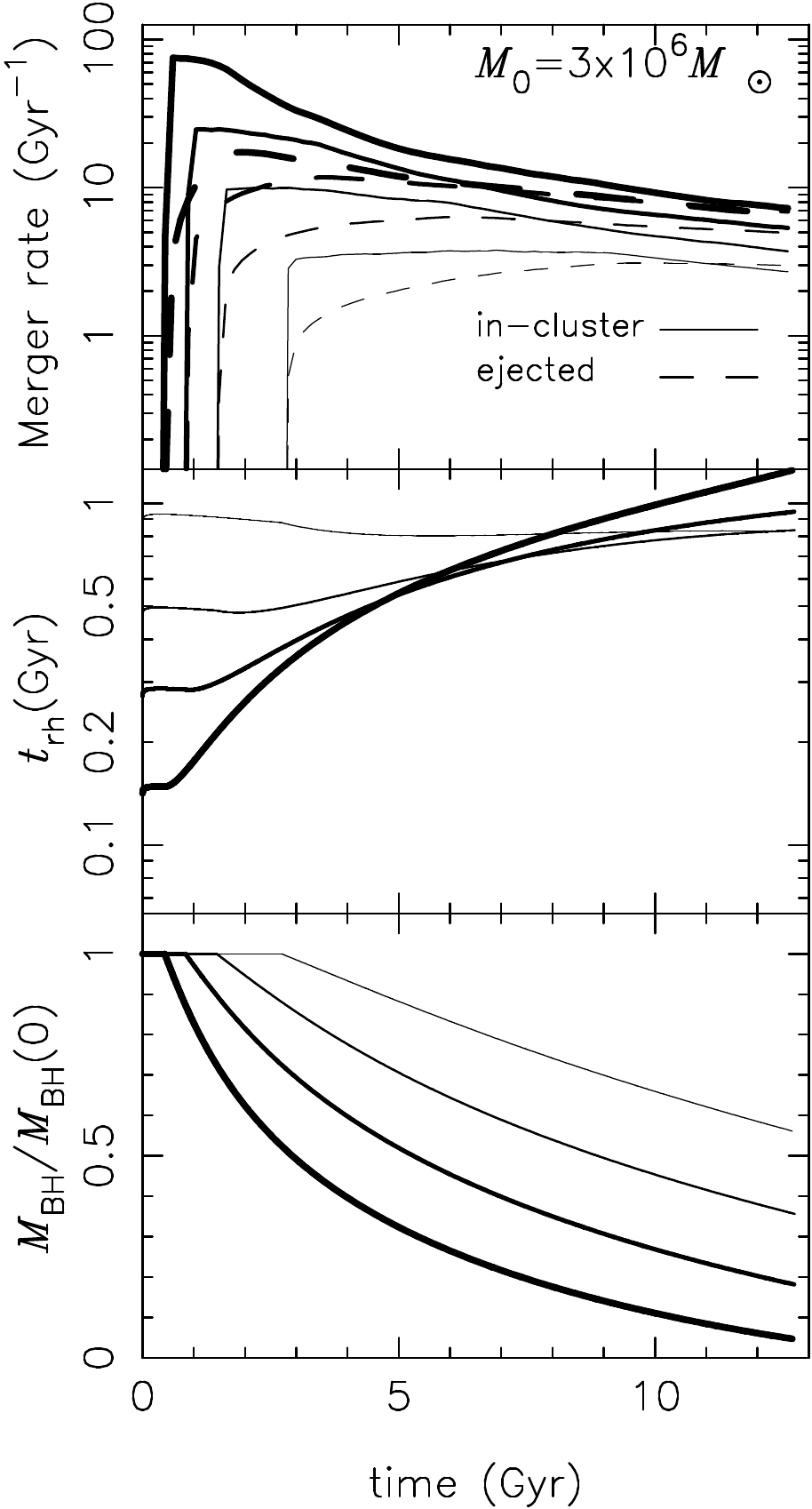}
     \includegraphics[width=1.58in,angle=0.]{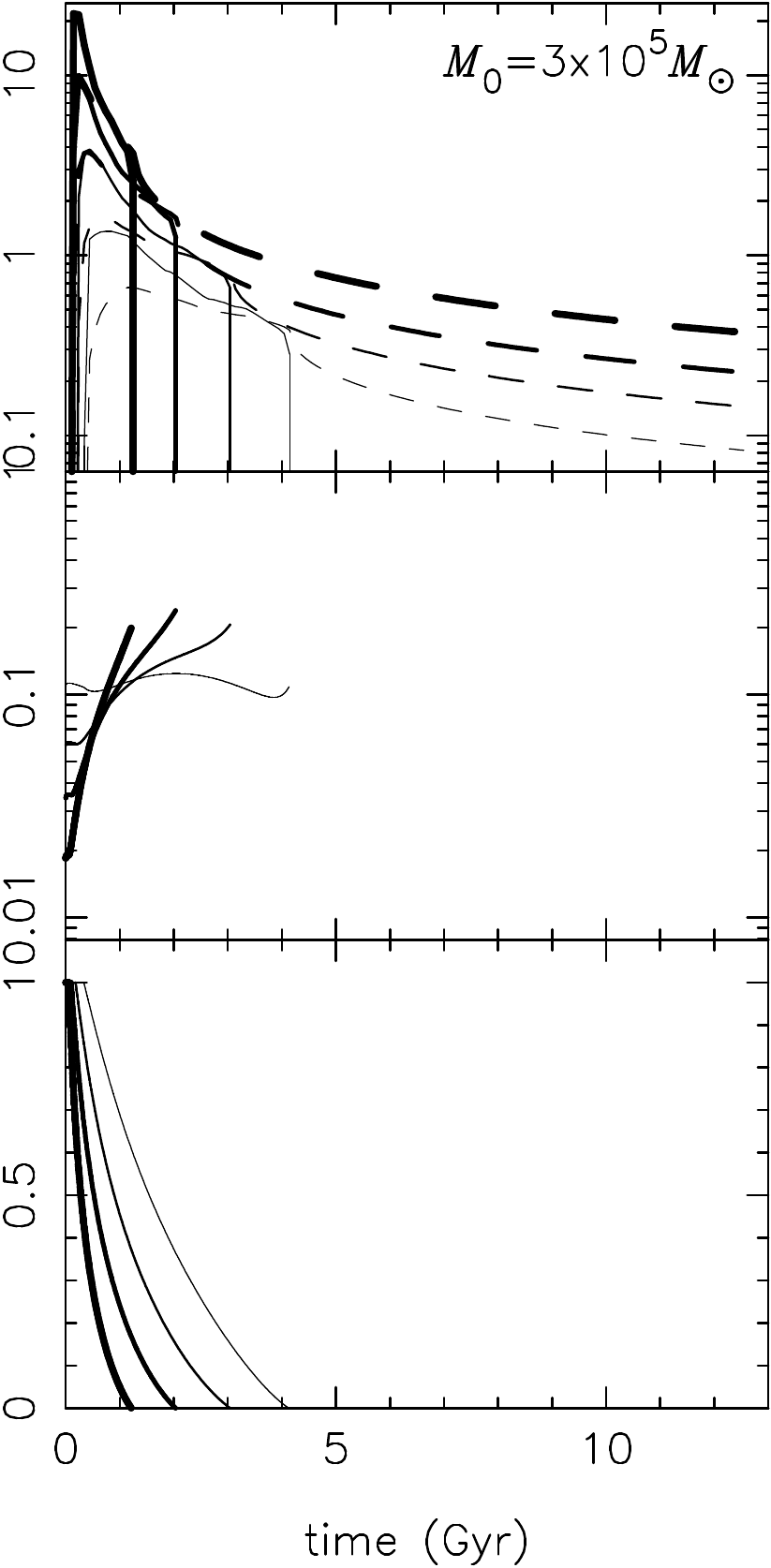}
     \caption{Example of the BHB merger rate evolution for
       two cluster masses, and separately for in-cluster  and
       ejected binaries. The middle panels gives the cluster relaxation time, and
       the lower panels the total BH mass in units of the initial value.
       Different lines correspond to different initial half-mass density,  $\rho_0=0.03,\ 0.1, 0.3$
       and $1\ \times 10^5 \mspc$. The initial density increases with line thickness. Here we set
       $\Delta=3\times 10^5\msun$ and the simulations are terminated either after $13\,$Gyr of evolution or
       after all BHs have been ejected.    } 
    \label{fig11}
\end{figure}

\begin{figure*}
    \centering
    \includegraphics[width=3.4in,angle=0.]{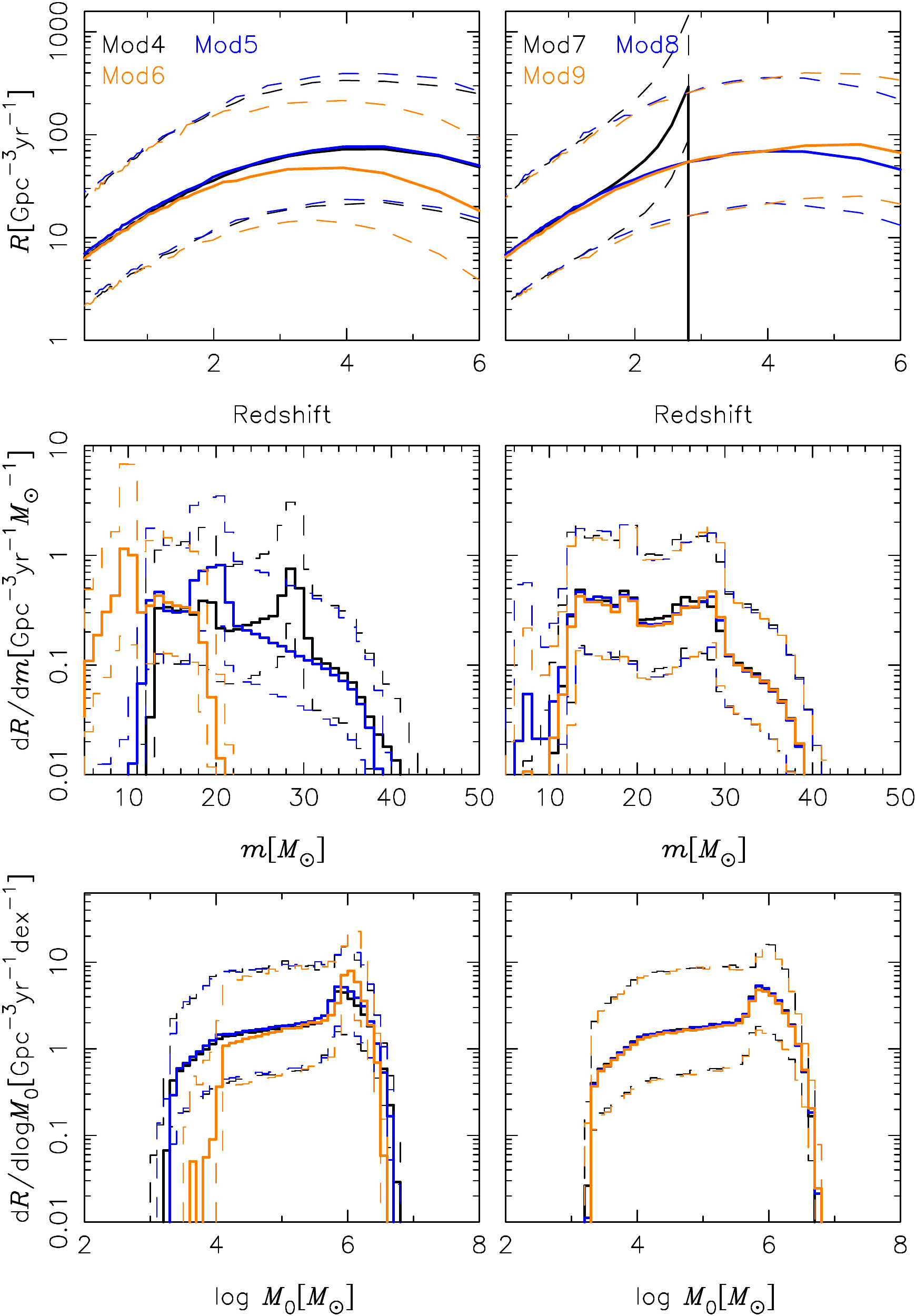}\includegraphics[width=3.12in,angle=0.]{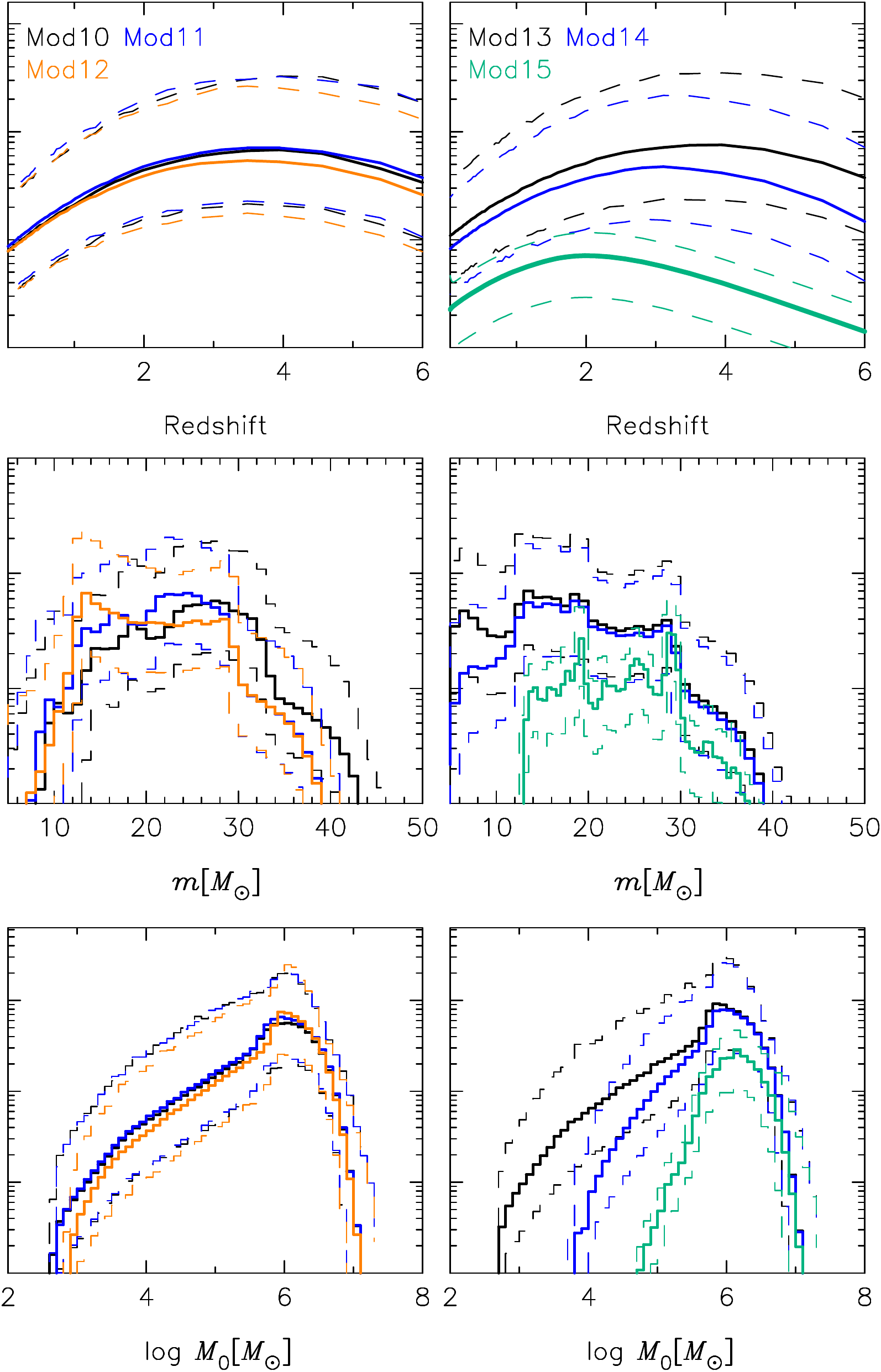}
    \caption{Median of the merger rate distribution (solid lines)
      and $90\%$ confidence intervals (dashed lines) for the models Mod4 to Mod15 in Table~\ref{T1}
      where the initial half-mass density is set to $10^4\mspc$.
      Middle panels give the  distribution of primary BH masses for mergers at $z<1$. The lower panels show the mass distribution
      of clusters where these merging binaries were formed.
    } 
    \label{figALL}
\end{figure*}

\begin{figure*}
    \centering
    \includegraphics[width=3.4in,angle=0.]{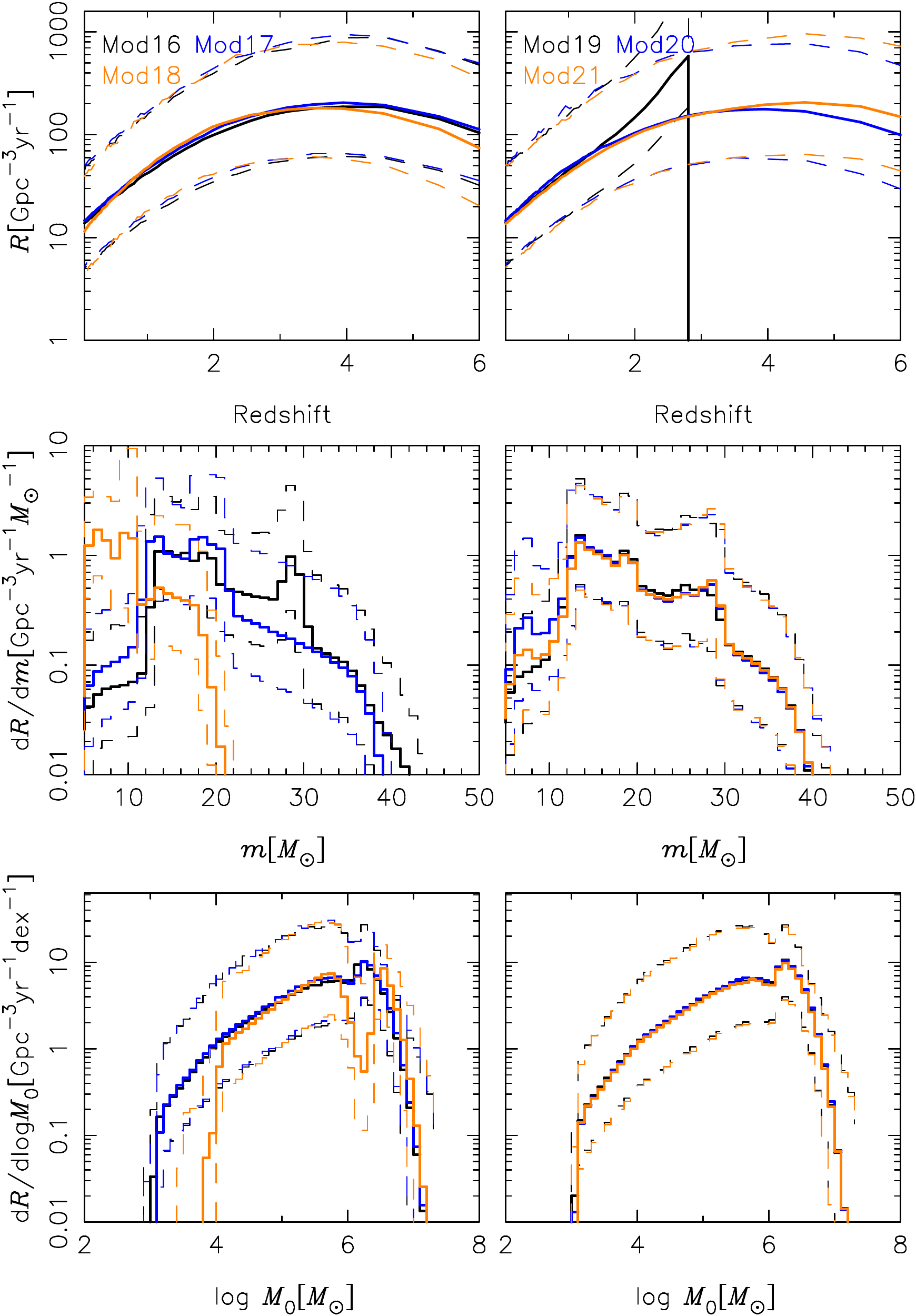}\includegraphics[width=3.12in,angle=0.]{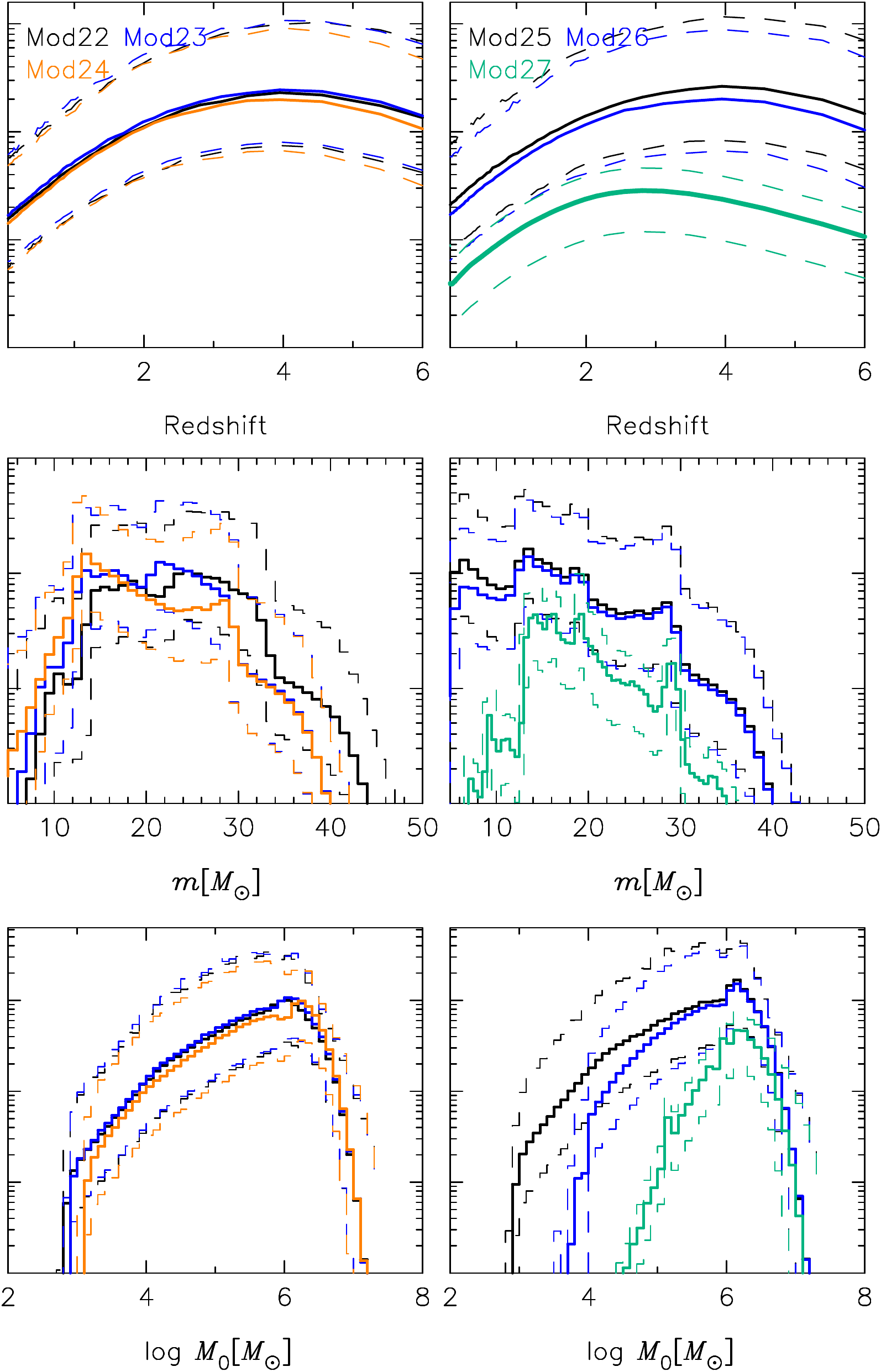}
    \caption{Same as Fig.~\ref{figALL} but for $\rho_0=10^5\mspc$, i.e., Mod16 to Mod27 in Table~\ref{T1}.
    } 
    \label{figALL2}
\end{figure*}

\begin{figure*}
    \centering
    \includegraphics[width=3.4in,angle=0.]{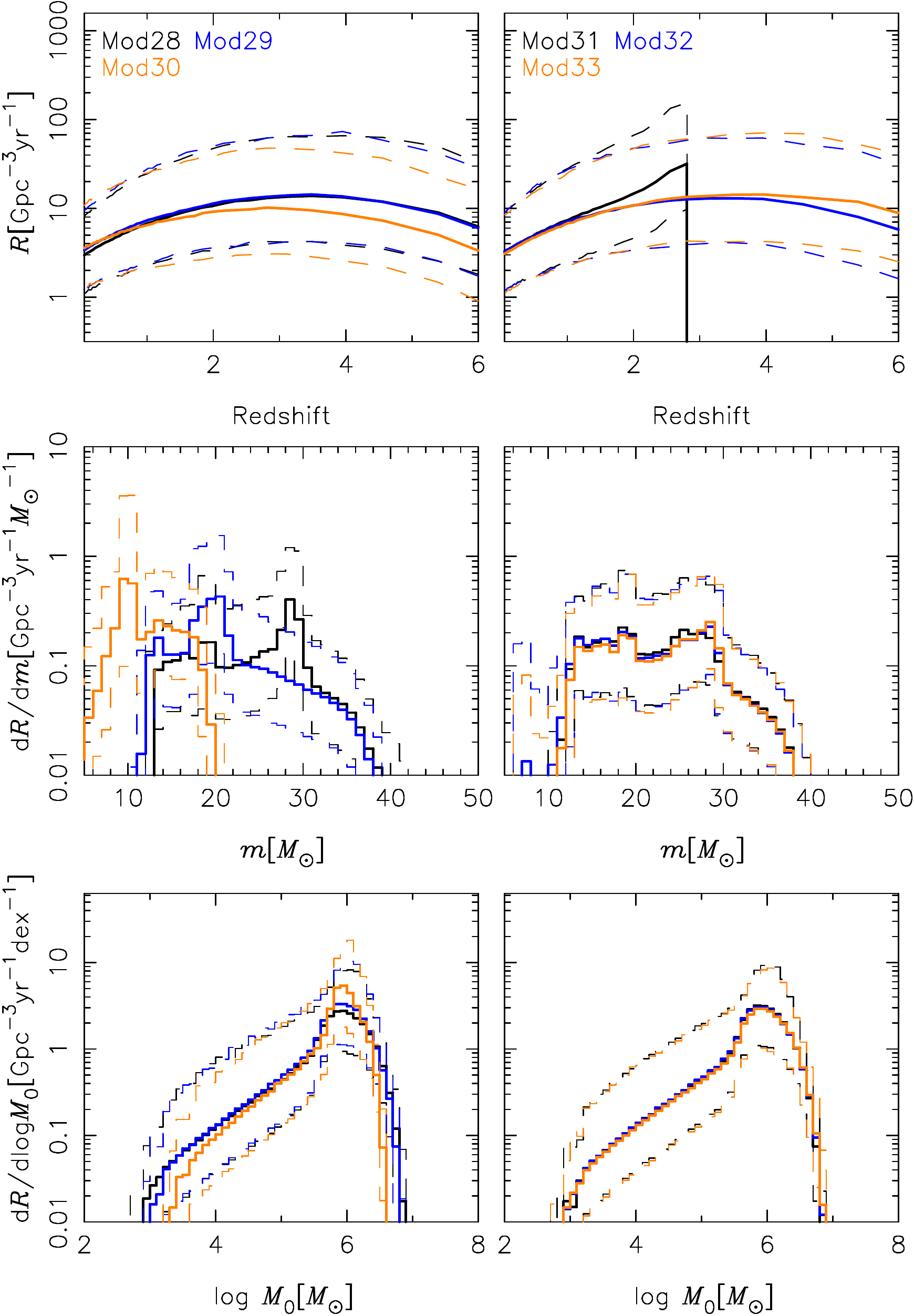}\includegraphics[width=3.12in,angle=0.]{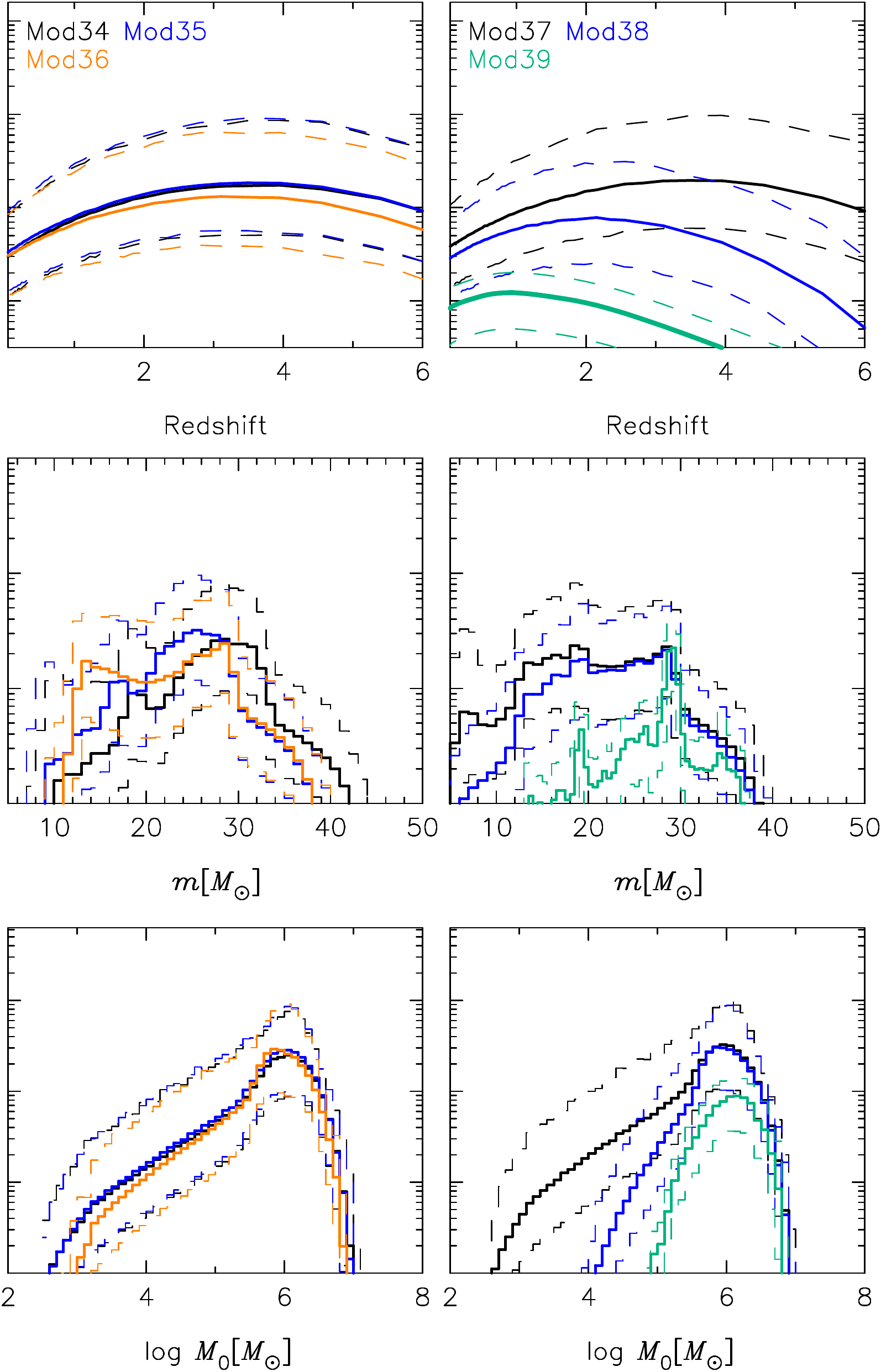}
    \caption{Same as Fig.~\ref{figALL} but for $\rho_0=10^3\mspc$, i.e., Mod28 to Mod39 in Table~\ref{T1}.
    } 
    \label{figALL3}
\end{figure*}

Another result of our analysis  is that the local rate of in-cluster inspirals
and GW captures are nearly independent
of the initial density assumed.
In general, we
find that other model variations also have little effect on the merger rate of in-cluster binaries.
This result  is  because cluster evolution during balanced evolution, { i.e. at late stages when in-cluster mergers that we can observe occur, is insensitive to the initial conditions}.

Due to the expansion powered by the BHBs, all clusters evolve  asymptotically to
(approximately) approach the same value of half-mass relaxation time.
Hence, after some  time, the merger rate of in-cluster binaries must also become approximately the same for all clusters. This concept is  illustrated in Fig.~\ref{fig11} where  we show the evolution
of a set of cluster models with the same  mass but different $\rho_0$. 
For $M_0=3\times 10^6\msun$ (left panels), the BHB merger rate at $\gtrsim 10$\,Gyr only varies by a factor of
$\sim 2$ between the models, although  these were started with widely different densities.
The middle panel gives the cluster half-mass relaxation time, which also tends to evolve
to the same value for all models.  This  roughly recovers H\'{e}non's result that
$t/t_{\rm rh}$ increases until $t_{\rm rh}\sim t_{\rm rh,0}/\zeta $, after which $t_{\rm rh}\propto t$ \citep{2010MNRAS.408L..16G}.

In the right panel of  Fig.~\ref{fig11} we consider the evolution of clusters with an initial lower mass $M_0=3\times 10^5\msun$. In these 
 models all BHs are ejected at $t\lesssim 5$\,Gyr.
Thus, their  in-cluster binaries do not contribute to the merging population at late times.
This explains why in the population models above the
local merger rate of in-cluster binaries is widely dominated by high mass systems,  
$M_0\gtrsim 10^6\msun$ (e.g., lower panel of Fig. \ref{figin}).
Moreover, the binary merger rate near the end of the simulations shows a larger variation among 
different models than in the high mass cluster case.
This simply reflects the large difference in the cluster density at early times
when these binaries were formed and ejected.

\subsection{Dependence on  model parameters}\label{Mpar}
In the previous section we  consider the merger rate density evolution 
for three  choices of initial cluster density. Here we discuss the results for a larger set of models
in which for each of the three density assumptions we 
vary the prescription for the cluster metallicity, the cosmological cluster formation  model, 
the  BH formation mechanism,  the BH natal kicks, and the cluster mass loss rate.
All the model parameters we considered  are listed in Table~\ref{T1} (Mod4 to Mod39), together with the corresponding
values of merger rate evolution parameters and uncertainties.
We stress that not all the models analysed here are realistic representations of a globular cluster population. 
They are nevertheless useful in order to understand the impact of different model parameters
on the merger rate evolution and BH mass distribution.
 Our main message
 here is that variations in other model assumptions have little effect on the local value of the BHB merger rate density and its redshift evolution. Thus, we conclude this section by 
 presenting the results from an additional set of models where $\rho_0$ is varied
over a wider  range of values than in Section~\ref{MR} to more systematically explore its effect on  the  BHB merger rate.

{\it Metallicity}.  In order to explore the
 dependence of the merger rate on metallicity, we consider models where the clusters all have  the same metallicity which we
 set to $Z=0.01$, 0.1 or $1\times Z_\odot$.
 Since mass loss due to stellar
winds is less effective in metal-poor stars, the  forming merger remnant mass increases with decreasing  metallicity.
 At Solar metallicity, the mass distribution of the final merger products spans from a few solar masses up to about
 $30 \,M_\odot$ and peaks near $10\,M_\odot$. At lower metallicity, $Z=0.01$ and 0.1, the distribution of remnant masses is much wider with its maximum at $\sim 50M_\odot$.
 This has an obvious effect on the  mass distribution of the  merging BHBs  
 as can be seen in  Fig.~\ref{figALL}, \ref{figALL2} and \ref{figALL3}. The value of metallicity affects also what type of clusters make the BH mergers in the local universe,
 with their mass distribution being skewed towards higher values for Solar metallicities.
  The important result  here, however, is that the evolution of the merger rate density is largely
 unaffected by the choice of metallicity and its dependence on cluster age. Even in the unrealistic case in which all clusters are formed at Solar metallicity,  
  the merger rate density only starts to deviate significantly from  the other models at $z>2$.
 Such lower merger rate at early times
 is expected and it is a consequence of the longer $t_{\rm rh}$  due to  the
 lower initial BH mass fraction.

 We conclude that a detailed knowledge of the metallicity distribution of GCs and its dependence
 on time is not necessary in order to determine a  BHB merger rate, although it has an important effect on their
 mass distribution.
 
{\it Cluster ages.}
  We implemented two additional  choices 
  for the  parameters in the cosmological model of \citet{2019MNRAS.482.4528E} which determine the distribution of cluster ages:
  $[\beta_\Gamma=1,\ \beta_\eta=1/6]$, and $[\beta_\Gamma=0, \ \beta_\eta=1/3$].
These two models are shown in Fig. 8 of \citet{2019MNRAS.482.4528E}.
Moreover, we consider an additional  case of a burst-like cluster formation history in which all clusters are formed at $z=3$.

From Fig.~\ref{figALL}, \ref{figALL2}, and \ref{figALL3} we can see that, for a given initial density, our results at $z\lesssim2$ are also  independent of
the exact distribution of cluster formation times. Within this redshift, even the oversimplified case in which all clusters
form at $z=3$ leads to a merger rate density and BH mass distribution that are consistent with those obtained from the full cosmological models.
At $z>2$, however, the redshift evolution of the merger rate is clearly affected with its peak coinciding with the peak of cluster formation activity in each model.

{\it BH formation.}
We consider three more recipes to computing
the  BH mass distribution  based on different core-collapse/supernova models.
We use the {\it delayed}
model in which the supernova explosion is allowed to occur over a much longer timescale than in the previously employed rapid model \citep{Fryer2012}.
We then use the  compact-object mass prescriptions from  \cite{Belczynski2002} and \cite{Belczynski2008}. These two latter models use slightly different
recipes for the
proto-compact object masses while adopting the same formulae to determine the amount of fallback material.
We note that the effect of the BH formation recipe is two folds as it influences both the  mass distribution of the BHs as well as
their natal kicks. 
Apart from the effect on the BH mass function, however, there is very little change of the  merger rate evolution among the various prescriptions,
with the delayed model leading to a slightly lower merger rate at all redshifts than the others.

{\it Natal kicks.}
Two additional assumptions about  the BH  natal kicks  are explored.
In one the BHs are formed with no kick, and  in the other the BHs receive the same momentum kick as neutron stars,
meaning that their kick velocities  are drawn from a Maxwellian distribution with dispersion
$\sigma=265\,\kms$ \citep{Hobbs2005}
and then reduced by the neutron star to BH mass ratio, $1.4M_\odot/m$.
Among the model variations considered in this section, the BH natal kick prescription has the largest (but still mild) impact on our results.

The zero kick and the momentum kick prescriptions lead, respectively, to a larger and smaller retention fraction of BHs compared to the fallback prescription \citep{2019arXiv190207718B}.
The difference becomes especially important in clusters with initial mass $M_0\lesssim 10^4M_\odot$  because of their lower escape velocities. In these clusters, virtually no BHs
are left  after the momentum kicks have been imparted, which is  reflected in the mass distribution of useful clusters shown in the bottom-right panels of Fig.~\ref{figALL}, \ref{figALL2}, and \ref{figALL3}.

{\it Cluster evaporation.}
Our mass-independent {and orbit-independent mass loss rate for} cluster evaporation is certainly a simplified one.
To understand its effect on the cluster and BHB evolution, we
computed three additional models with exactly the same initial conditions as in Mod1, Mod2 and Mod3 but with $\dot{M}_{\star,\rm ev}=0$.
Here we still compute the initial GCMF from equation~(\ref{CIMF}) and use the $[\Mc,\Delta]$ values obtained from the MCMC analysis above, but we do not include any prescription for mass loss when evolving the clusters. 
Thus, this exercise is only meant to determine the importance of the mass loss effect on the secular evolution of the clusters and the BHBs they produce.
{ We find that in these new models,  the local value of the merger rate density and of $\kappa$, as well as the BH mass and progenitor cluster mass distributions are consistent with those found in 
the  models with cluster evaporation included. }
For  the same initial conditions as in Mod1, Mod2 and Mod3
the median values of the local merger rate are $R_0=6.9\, \rm Gpc^{-3} yr^{-1}$, $14.1\, \rm Gpc^{-3} yr^{-1}$ and $3.5\, \rm Gpc^{-3} yr^{-1}$, respectively.
This shows that cluster evaporation has a small effect on the dynamics of the BHBs.

In our models, however, tidal mass loss must become important at some point, e.g., for high enough $\Delta$,  GCs will evaporate before they can produce BHBs.
We now quantify how high $\Delta$ needs to be in order to change the BH dynamics significantly.
To do this we compare the tidal mass loss timescale, $t_{\rm ev}\equiv M_0/|\dot{M}_{\star,\rm ev}|$,  to the timescale
after which the BHs have been nearly depleted by dynamical ejections,
which we define to be $t_{\rm BH}\equiv M_{\rm BH,0}/|\dot{M}_{\rm BH}|$. We should expect that
for $t_{\rm BH}< t_{\rm ev}$ most BHBs will have formed already  before the cluster mass
has changed significantly  due to evaporation. This will happen if $\Delta$ is smaller than the critical value
\begin{equation} 
\Delta_{\rm c}\simeq {\langle t \rangle \over t_{\rm rh,0}} {\beta\over f_{\rm BH}}M_0\ .
\end{equation}
For  $\rho_0=10^3 \mspc$ and $ f_{\rm BH}=0.05$, we find $\Delta_{\rm c}\approx 10^6M_\odot$ independent of the initial cluster mass;
for  $\rho_0=10^5 \mspc$, we have  $\Delta_{\rm c}\approx 10^7M_\odot$.
These values are larger than any value of $\Delta$ used in our models (see Fig.~\ref{fig1}),
explaining the small impact of cluster evaporation on the results.

While the  
models discussed  above show that the impact of cluster evaporation on the BHB dynamics is small, they do not
asses its effect on the merger rate. Thus, we consider 
three new models with  $\dot{M}_{\star,\rm ev}=0$  but now use an initial GCMF that only accounts for mass loss due to stellar evolution. 
If only stellar evolution is included, the initial GCMF { that gives rise to the present-day GCMF shown in Fig.~\ref{fig1}} becomes: 
\begin{equation}\label{psip}
\phicln^\prime=0.5A(M_0/2+\Delta)^{-2} \exp\left[-(M_0/2+\Delta)/\Mc\right]
~ ,
\end{equation}
and $K\simeq 2$.
These new models provide us with a  safe lower limit on the BHB merger rate for each density assumption; they are Mod15, Mod27 and Mod39 in Table~\ref{T1} and Fig.~\ref{figALL}, \ref{figALL2} and \ref{figALL3}.
From these results we see that the merger
rate in models without evaporation  are about three times  smaller than in models where the effect of cluster evaporation is included.

\begin{figure}
    \centering
    \includegraphics[width=3.4in,angle=0.]{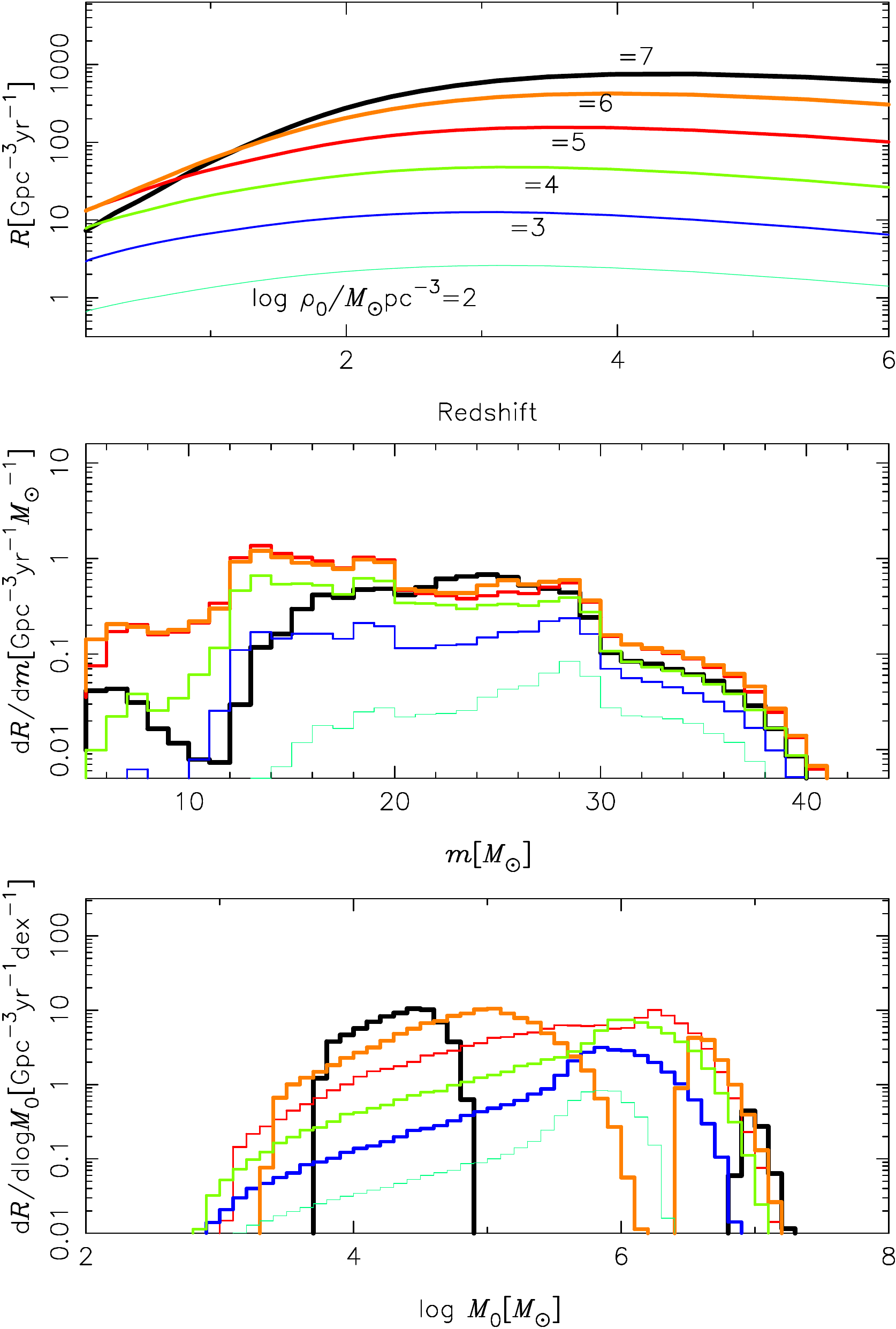}
    \caption{ The upper panel shows the median of the merger  rate density
    distribution as a function of redshift;  the middle panel gives
    the distribution of primary BH mass for $z<1$ mergers (median values); and  the bottom panel gives the initial  mass distribution of clusters contributing to the local mergers (median values). In these calculations we varied the initial cluster density within the indicated range while keeping all the other model parameters the same and as in Mod1 of Table~\ref{T1}.
    } 
    \label{RHOdep}
\end{figure}
 
\begin{figure*}
    \centering
    \includegraphics[width=3.2in,angle=270.]{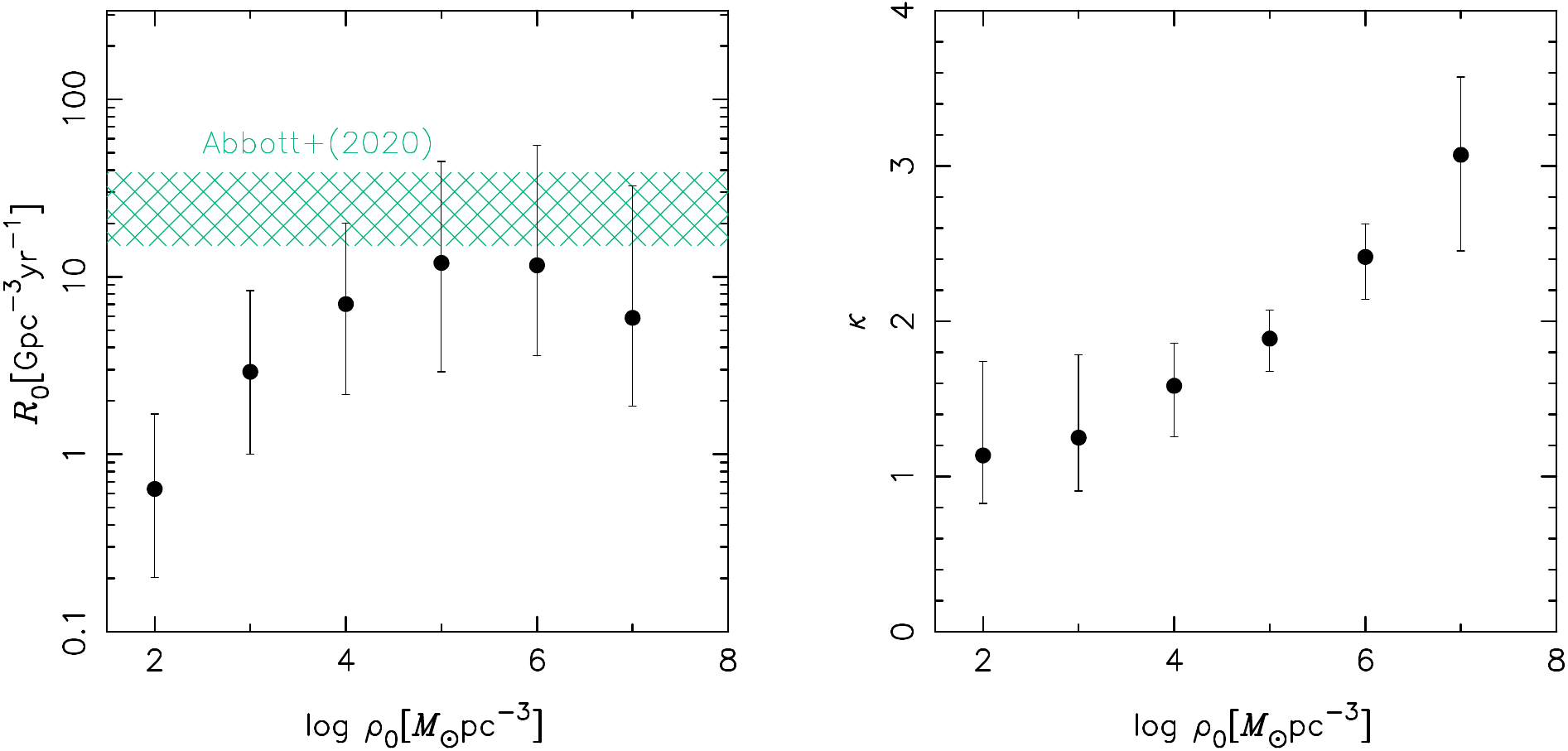}
    \caption{Merger rate parameters as a function of the initial cluster half-mass  density for the models of Fig.~\ref{RHOdep}.
    The black points represent median values, while the lower and upper error bars give the 5 and  95  percentiles  of the distributions.      
    } 
    \label{RHO}
\end{figure*}

\begin{figure}
    \centering
    \includegraphics[width=3.4in,angle=0.]{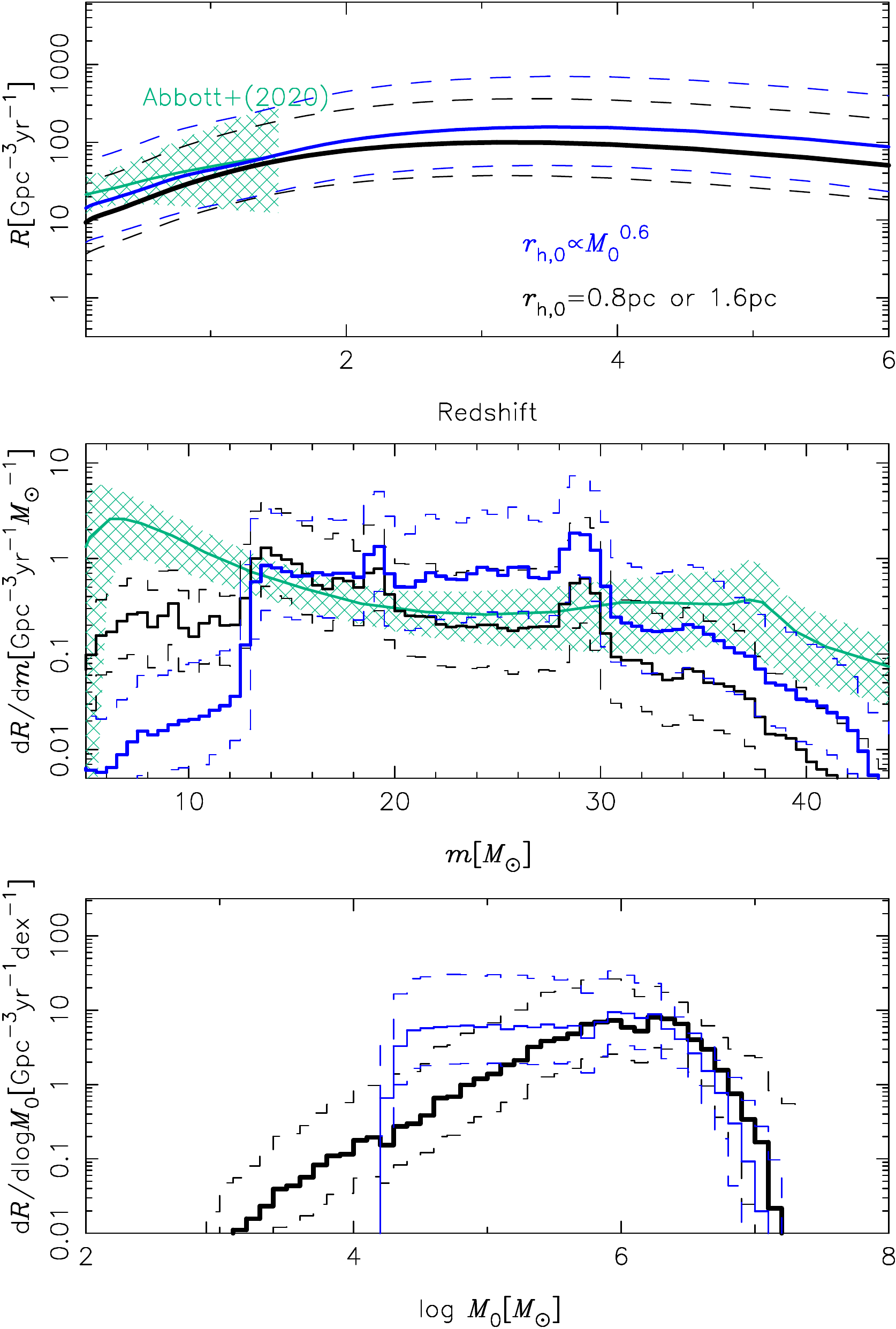}
    \caption{Merger rate evolution, primary BH mass of mergers at $z<1$ and the initial cluster mass distributions where these binaries originated, for a model where half of the clusters
    have $r_{\rm h,0}=$0.8\,pc and the other half have  $r_{\rm h,0}$=1.6pc similar to  \cite{2018ApJ...866L...5R} (black lines),
    and for a model where $r_{\rm h,0}\propto M_0^{0.6}$ (blue lines).
    } 
    \label{fig12}
\end{figure}

{\it Cluster density.}
The model variations explored above show that for a given initial GCMF, the initial cluster density is clearly the most
important parameter for setting the BHB merger rate density and its redshift evolution. Thus, here we perform
 a  more systematic exploration of such dependence by running an additional set of models where the initial cluster density is varied in the range $\rho_0=10^2\mspc$ to $10^7\mspc$. All other model parameters are set as in Mod1 of Table~\ref{T1}. 
The results from these additional models are shown in Fig.~\ref{RHOdep} and Fig.~\ref{RHO}.

From Fig.~\ref{RHOdep}  we see that  the peak of the merger rate  and the redshift at which it occurs
in each model increase with $\rho_0$, and vary from $\simeq 3\, \rm Gpc^{-3}yr^{-1}$ at
$z=3$
for  $\rho_0=10^2\,\mspc$, to  $\simeq10^3\, \rm Gpc^{-3}yr^{-1}$  at $z=4.5$ for  $\rho_0=10^7\,\mspc$. 
The situation is different, however, when we look at the merger rate in the local Universe.
In Fig.~\ref{RHO} we see that the median value of $R_0$
has a maximum value of $\simeq 20\,\rm Gpc^{-3}yr^{-1}$ at  $\rho_0\simeq10^5\,\mspc$.
This is an important result as it shows that the local BHB merger rate density from the GC channel
has a robust upper limit of
$\simeq 50 \,\rm Gpc^{-3}yr^{-1}$ --  the upper error bar estimate  for the $\rho_0=10^5\mspc$
model.

The reason why  $R_0$ decreases with $\rho_0$ above a certain density can
  be understood from the lower panel in  Fig.~\ref{RHOdep}. This plot  shows the
mass distribution of  clusters from which the BHBs  that merge in the local universe are formed.
For initial densities above $10^5\mspc$ the contribution from clusters in the mass range  $10^5M_\odot\lesssim M_{0}\lesssim 10^7M_\odot$
gradually decreases as the {distribution of cluster masses contributing to the mergers} becomes bimodal.
Such narrowing of the range of cluster masses that can produce local mergers explains  the 
 relatively low merger rate in the higher density models. It also
 affects the distribution of the primary BH masses as shown in the middle panel of Fig.~\ref{RHOdep}.

We looked into the density dependence of the cluster mass distribution shown in the lower panel of Fig.~\ref{RHOdep} in more details. We find that the lower mass peak seen 
 in the cluster mass distribution
 for $\rho_0=10^6\mspc$ and  $10^7\mspc$ 
 is  only due to ejected binaries while
 the  higher mass peak is only due to in-cluster mergers. 
 Thus, we can explain the depletion of BHBs that come from intermediate mass clusters by considering the behaviour of the two merging populations
when varying  $\rho_0$.
Above a certain initial cluster mass, $v_{\rm esc}$ becomes large enough ($\gtrsim 100\kms$) that  all BHBs merge inside the cluster.
But, because $v_{\rm esc}\propto M^{1/3}\rho^{1/6}$, the value of initial cluster mass that still allows for dynamical ejections to occur
goes down as $\rho_0$ increases.  This explains why the upper end of the mass distribution of clusters that produce mergers from ejected 
BHBs decreases as $\rho_0$ increases. Clusters with an initial mass larger than this value only produce in-cluster mergers.
Such clusters, however, can only produce mergers in the local universe if they still contain BHs at the present time.
Because  the BHs are processed at a rate $t_{\rm BH}^{-1}\propto \sqrt{\rho}/M$, the value of the initial cluster mass above which BHs
are still present in the local universe increases with density. 
This explains why the lower end of the mass distribution of systems that produce in-cluster mergers moves towards larger masses
as  $\rho_0$ increases;  clusters with a mass smaller than this value get rid of all their BHs by $z=1$. 

Fig.~\ref{RHOdep} shows that only
models in which GCs start with an initially high density $\rho_0\gtrsim 10^4\mspc$ can account for a large fraction of
the LIGO-Virgo BHB mergers. Interestingly, these models also give a better fit to the 
inferred BH mass function above $m\gtrsim 13M_\odot$ as seen in Fig~\ref{fig2}.
Future GW observations will reduce the error bars 
associated with the merger rate estimates and the BH mass distribution,  providing important clues on the  the initial densities of GCs.

Finally, we consider two additional model realizations. In one we 
 evolve the same initial conditions as in
 \cite{2018ApJ...866L...5R} and \cite{2018PhRvD..98l3005R}
 where half of the clusters have
 $r_{\rm h,0}=0.8$\,pc and the remaining half have $r_{\rm h,0}=$1.6\,pc;  in the other model,    the cluster half-mass radius scales as
\begin{equation}\label{mrr}
  \log\left({{r}_{\rm h,0} \over {\rm pc} }\right)=-3.56+0.615 \log \left( {M_0\over M_\odot }\right) \ .
\end{equation}
This latter relation  {was derived by \cite{2010MNRAS.408L..16G} from the results of}  \citet{2005ApJ...627..203H} {who fit this Faber-Jackson-like relation  to  ultra-compact dwarf galaxies (UCDs) and elliptical galaxies. \citet{2010MNRAS.408L..16G} derived the initial mass-radius relation correcting for mass loss and expansion by stellar evolution and correcting radii for projection.}  All the other model parameters are the same as in Mod1 of Table~\ref{T1}.
 The results of these two new models are shown in Fig.~\ref{fig12}. Interestingly, both   give a local merger rate, $\simeq 10 \,\rm Gpc^{-3}yr^{-1}$, which is similar to the maximum merger rate value we obtained before. Moreover,
these results show how the choice of initial half-mass radius relation has a significant effect on both the BH mass  and initial cluster mass distributions. For $r_{\rm h,0}\propto M_0^{0.6}$, the cluster mass distribution becomes nearly flat so that, roughly speaking, all clusters with initial mass in the range $10^4-10^6M_\odot$ contribute equally to the local merger rate.

 \section{Discussion and conclusions}
\label{sec:discussion}

Our models are based on some  assumptions and  approximations. These are discussed in the following  sections, which also present a more detailed comparison to the literature. We end the paper with a brief summary of our main results.

\subsection{Present-day GC mass density}
 Our  value of $\rhogc$ is larger than what was found by \cite{Rodriguez2015a}: if we adopt their assumption of an average GC mass of $3\times10^5~\msun$, we find that equation~(\ref{eq:rhogc}) corresponds to a GC number density of $\ngc=2.4~\mpc^{-3}$, which is $3.3$ times larger than their  $\ngc=0.72~\mpc^{-3}$, but similar to the value used in {\cite{2000ApJ...528L..17P} and} \cite{2020ApJS..247...48K}. Part of this difference is because we adopted a larger value of $\eta$: if we use their mild $\Mh$-dependent $\eta$ from \cite{2015ApJ...806...36H},  we find $\ngc=1.50~\mpc^{-3}$, which corresponds to our lower error bar $\langle\eta\rangle$. However, this is still a factor of 2.1 higher than what was found by \cite{Rodriguez2015a}. We are not sure what causes this remaining difference, but we note that $\ngc=1.50~\mpc^{-3}$ is about a factor $h^{-2}$ larger than $\ngc=0.72~\mpc^{-3}$ {(Carl Rodriguez, private communication)}. 
\subsection{Initial GC density in the Universe}
To derive $\rhogcn$, a different approach was adopted by  \cite{2018ApJ...866L...5R} and \cite{2018PhRvD..98l3005R}. They use the total mass density of GCs forming in the semi-analytical galaxy formation model of \citet{2019MNRAS.482.4528E}. They approximate the numerical results with analytical functions and find a total $\rhogcn= 5.8\times10^{14}~\msun\,\gpc^{-3}$, about 15\% higher than \citet{2019MNRAS.482.4528E} and 20\% lower than our adopted $\rhogc$ (equation~\ref{eq:rhogc}). They then assume that the initial masses of all GCs were a factor of 2.6 higher (from \cite{2019ApJ...873..100C}) because of stellar mass loss and evaporation and find that initial mass density of GCs more massive than $10^5~\msun$ is $\rhogcn(M_0>10^5~\msun)\simeq1.5\times 10^{15}~\msun\,\gpc^{-3}$.
This is a factor of $\sim4$ higher than found by  \cite{2018ApJ...866L...5R}. 
 The reason we find a higher value is that their assumption that the present-day GC density in the Universe is made from GCs with $M_0>10^5~\msun$ that lost (only) a factor of 2.6 in mass implies a mass loss rate that is much lower in our models. In our models the present-day $\rhogcn$ is made of GCs with $M_0\gtrsim4\times10^5~\msun$. 
 
 In addition, we do consider the contribution to the merger rate of lower mass GCs  with $M_0<10^5~\msun$. 
We Use \modelbbr\
 to evolve the same initial conditions as in
 \cite{2018ApJ...866L...5R} and \cite{2018PhRvD..98l3005R}
 where half of the clusters have
 $r_{\rm h,0}=0.8$\,pc and the remaining half have $r_{\rm h,0}=$1.6pc but extended the initial GCMF down to $\Mlo=100M_\odot$ (see Fig. \ref{fig12}). We find that 
 $\simeq 10\%$ of the local mergers come from GCs with $M_0<10^5~\msun$, and
 therefore conclude that for the exact same initial GCMF, our models would lead to a local merger rate that is still $\simeq 4$ times that found in  \cite{2018ApJ...866L...5R} and \cite{2018PhRvD..98l3005R}. 
 Our Schechter mass is $\Mc\simeq 2\times10^6M_\odot$, so we compare to the results from \cite{2018ApJ...866L...5R} for   mass functions with $\Mc = (2.5-5)\times10^6 M_\odot$. For these, the rates in  \cite{2018ApJ...866L...5R} are   $5$ to $10\rm Gpc^{-3}yr^{-1}$. 
 The median of the merger rate distribution computed from our  models
 is  $10\rm Gpc^{-3}yr^{-1}$.
 \citet{2018ApJ...866L...5R} use a fit to the results of a set of Monte Carlo simulations to determine the number of mergers produced by a cluster as a function of time. Because these fitting formulae are not public, it is currently difficult to establish the reason   why our rates are only 1 to 2 times, and not 4 times,  those in \cite{2018ApJ...866L...5R}. We note in passing that  we compared our models to the number of mergers from the two examples shown in Fig. 2B of \cite{2018ApJ...866L...5R} and found very good agreement.

\subsection{O-star ejections and IMBH formation}
\label{ssec:oimbh}
\begin{figure}[!h]
    \centering
    \includegraphics[width=3.4in,angle=0.]{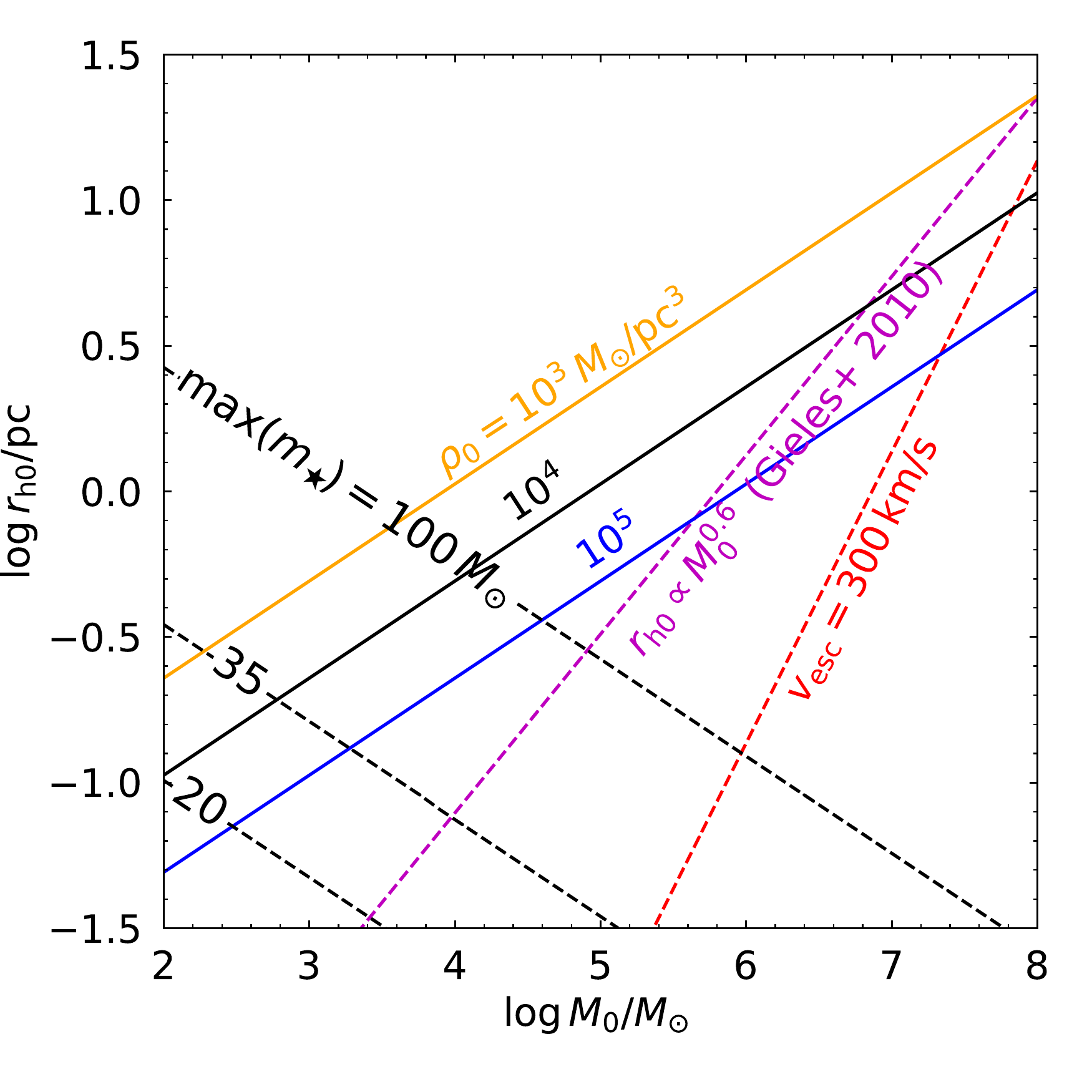}
    \caption{GC initial mass-radius diagram showing the 3 initial cluster densities adopted in this work with dashed lines. The magenta, full line shows the initial mass-radius relation that describes the most massive GCs ($\gtrsim10^6\,\msun$) and ultra-compact dwarf galaxies ($\gtrsim10^7\,\msun$) \cite{2005ApJ...627..203H,2010MNRAS.408L..16G}. Full lines show the maximum mass of O-stars that form BHs inside the cluster.  Clusters with $M\lesssim10^4\,\msun$ will eject some O-stars before they become BHs, and these clusters will therefore have slightly lower BHB mergers than in our model. { The red, full line shows an initial escape velocity of 300\,km/s, which is the minimum escape velocity required for IMBH formation to occur \cite{2019MNRAS.486.5008A}}. This process is not playing a role in our adopted initial conditions. } 
    \label{fig:mr}
\end{figure}

{Our  \modelbbr\ model makes the simplifying assumption that all BHs are in place when the cluster forms. Because the typical timescale of GC evolution (i.e. 100 Myr - Gyr) is much longer  than the timescale of BH formation (i.e. 10 Myr), this is fine in most cases. However, for very dense, low-mass clusters, the relaxation time is so short that O-stars are ejected as `runaway stars' before they form  BHs \cite{1967BOTT....4...86P,2011Sci...334.1380F}. 
As a result,   the  initial BH fraction is lower in these clusters than what we assume in our models,  possibly affecting the merger rate and properties of the mergers. To quantify this, we use the fact that the dynamical process that ejects O-stars is the same as the one that ejects BHs at a later stage. We therefore adopt \model\ and replace the BH population by a massive star population between $10-100\,M_\odot$, with a logarithmic slope of $-2.3$, and a mass fraction of 18\%, as appropriate for a Kroupa IMF. We then determine for a grid of initial cluster masses and half-mass radii the maximum mass of  massive stars that form BHs inside the cluster. In Fig.~\ref{fig:mr} we show contours for $20\,M_\odot$ 
(the minimum mass of an O-star to produce a BH), $35\,M_\odot$ (approximately half of the mass in BHs is produced by stars more massive than this) and $100\,M_\odot$ (the upper limit of our IMF). We also overplot the 3 initial cluster densities adopted in the previous section. From this we see that clusters with $M_0\lesssim10^4\,M_\odot$ are affected by O-star ejections, which affects about half of the mass in the initial GCMF.  However, these low-mass GCs are only responsible for $\sim15\%$ of the mergers. The fraction of clusters for which more than half of the BH mass is ejected is only a few per cent. Clusters that  produce runaways, will have fewer massive BHs, leading to a slightly higher merger rate of slightly less massive BHs. This effect is small, but would lead to a slightly steeper BH mass function especially for the densest models. However, we conclude that runaway stars do not significantly affect our results and that the effect is probably smaller than other uncertainties in our model. }

Another process that is not included in \modelbbr\ is repeated mergers of BHs. After a merger, the BH merger remnant receives a general relativistic momentum kick of several 100\,km/s, and if this is smaller than the escape velocity from the center of the cluster, it can be involved in subsequent mergers
 \cite{2016ApJ...831..187A,2018PhRvL.120o1101R,2019PhRvD.100d3027R} possibly forming an intermediate-mass BH (IMBH) \cite{2019MNRAS.486.5008A}. This can only occur for an initial escape velocity  $\gtrsim300\,$km/s and in Fig.~\ref{fig:mr} we show that only in our densest ($\gtrsim10^5\,\msun/{\rm pc}^3$), most massive ($\gtrsim10^7\,\msun$) models this could happen. Ignoring the effect of IMBH formation via dry BH mergers is therefore not affecting our results. 
 { Although the formation of
IMBHs through repeated mergers is unlikely, we note that   hierarchical mergers can still contribute to the BH merger rate. As discussed above,   hierarchical mergers represent only ten percent or less of the total number of BHB mergers expected from GCs  \citep{2018PhRvL.120o1101R,2019PhRvD.100d3027R}. Thus, they will not affect significantly our integrated merger rate estimates.  On the other hand,  second-generation mergers can produce  BHs with a mass higher than predicted by stellar evolution alone, broadening the BH mass distributions we derived and populating them above $\sim 40M_\odot$.
}

\subsection{Cluster mass loss and initial GCMF}
{Our models adopt a constant mass loss for all clusters of $\Delta \simeq2\times10^5\,\msun$. This is what is required to evolve the initial GCMF with a power-law slope of $-2$ at low-masses to the peaked GCMF of old GCs, but it is inconsistent with some studies of GC evolution. Firstly,  $N$-body simulations of tidally limited clusters show that $\dot{M}\propto M^{1/3}$ \cite{2003MNRAS.340..227B}, rather than $\dot{M}\propto M^0$. Including this mass dependence in $\dot{M}$ and maintaining the constraint that all GCs formed with the same Universal initial mass function, implies that clusters need to lose more mass for the turn-over in the GCMF to move to $2\times10^5\,\msun$ \cite{2009MNRAS.394.2113G}, resulting in a  twice as large value of $K\simeq 64$ \cite{2009ApJ...698L.158K} as we found for a mass independent $\dot{M}$. Secondly, the models of \cite{2003MNRAS.340..227B} show that $\Delta$ for a typical Milky Way GC is  smaller and depends on the apocenter and eccentricity of the Galactic orbit. For the median Galactocentric distance of Miky Way GCs ($\sim5\,$kpc) and an age of 10 Gyr, these models find $\Delta \simeq 4\times10^4\,\msun$, i.e. a factor of 5 smaller than what we assumed. These simulations considered the secular evolution of clusters in a static tidal field and  therefore underestimate  mass loss of clusters if additional disruption processes are important. For example, interactions with massive gas clouds in the early evolution (first Gyr)  can be disruptive  \cite{1958ApJ...127...17S, 2006MNRAS.371..793G}, have a similar mass dependence as relaxation driven evaporation in a static tidal field \cite{2016MNRAS.463L.103G}  and  lead to a turn over in the GCMF \cite{2010ApJ...712L.184E,2015MNRAS.454.1658K}. If this is the cause for the value of $\Delta$, than $|\dot{M}|$ is much higher in the early evolution, which would affect the resulting merger rate. Because the relaxation time decreases if the mass reduces, including this type of mass evolution will lead to a higher merger rate than in our models with an $\dot{M}$ that is constant in time. }{The models of \cite{2003MNRAS.340..227B} also do not contain BHs and it has been shown that retaining a BH population significantly increases the escape rate of stars \cite{2019MNRAS.487.2412G,2020MNRAS.491.2413W}. The BH population can increase $|\dot{M}|$ by an order magnitude (Gieles et al., in prep), especially towards the end of the evolution. This implies a relatively low(high) $|\dot{M}|$($\trh$) in the early evolution compared to our models, leading to a reduction of the merger rate. In addition, for $\dot{M}\propto M^\gamma$, with $\gamma<0$, the required $K$ to get the turn-over at the right mass is lower than for $\gamma=0$. If BHs are responsible for the value of $\Delta$, our merger rates could therefore also be slightly overestimated for this reason. However, we do not expect this effect to be important for dense clusters ($\gtrsim10^4\,\msun/{\rm pc}^3$) because their BHB mergers are produced when the clusters are still unaffected by the Galactic tides. We plan to include the effect of relaxation driven escape in a tidal field in a future version of \modelbbr\  to address this issue.  }

{ Finally, we have assumed that all GC masses are drawn from an initial GCMF that is constrained by the shape of the Milky Way GCs. Although the present-day GCMF is remarkably universal across galaxies, variations in the inferred $\Mc$ and $\Delta$ values of a factor of $\sim5$ are found across GC populations in galaxies in the Virgo cluster \cite{2007ApJS..171..101J}. 
Higher $\Mc$ and lower $\Delta$ values are found in brighter galaxies. Although variations in $\Mc$ and $\Delta$ are partially captured by the uncertainties in $\Mc$ and $\Delta$, this  accounts  only for up to a factor of $\sim2$. We may therefore under-populate the most massive clusters. }

\subsection{Primordial binaries}

The effect of binaries that form {in the star formation process and undergo stellar evolution } in the first stages of
cluster evolution has not been discussed so far. We argue here that  primordial binaries have a negligible effect on the merger rate and the distribution of the BH masses we derived.
{Because of H\'e{non} principle, the energy generation rate by binaries is determined by the relaxation process in the cluster as a whole. Whether dynamically active binaries form in three-body encounters from single BHs, or in encounters involving BHBs that formed from primordial binaries will therefore result in a central binary with the same properties. } However, primordial binaries might affect the initial BH mass function
  due to binary evolution processes. But, because BHB mergers from primordial binaries in GCs are a subdominant population at low redshifts \citep[see Fig. 2 in][]{Rodriguez2015a}, the effect on the local BHB mass distribution  is  also expected to be small.

 \subsection{Conclusions}

In this paper we have considered the dynamical formation of BHB mergers in GCs. 
Using our new population synthesis code
\modelbbr\ we have evolved a large number of models covering a much wider set of initial conditions than explored in the literature. 
This allowed us to place robust error bars on the merger rate and mass distributions of the merging BHBs. We find that the GC channel produces BHB mergers
in the local universe at a rate of $7.2^{+21.5}_{-5.5}\,\rm Gpc^{-3}yr^{-1}$, where the error bars are mostly set by the unknown initial GC mass function and initial cluster density.
By comparing to the merger rate inferred by LIGO-Virgo, our results imply that a model in which most of the detected  mergers  come from GCs is consistent with current constraints. This would require, however, that GCs  form with half-mass densities larger than $\gtrsim 10^4\mspc$, and suppression of other formation mechanisms.
All our models show a drop in the merger rate of binary with primary BH mass outside the range $\simeq13-30M_\odot$, for which there is no
evidence in the gravitational wave data. This might suggest that another mechanism is responsible for the production of these sources.

Our results have a number of implications for  the formation of BHB mergers and GCs. The dependence of the merger rate and BHB properties (e.g., eccentricity, mass) on the model parameters suggests that a direct comparison to the gravitational wave data will allow us to place  constraints on the initial conditions of GCs and their  evolution. 
Our models will also help to understand other uncertain parameters that control the formation of BHs and their natal kicks. While these latter parameters have little effect on the merger rate, they have a  significantly impact on the masses of the merging BHBs. Thus, 
useful constraints could be placed once the number of gravitational wave detections will be large enough to to allow for a statistically significant comparison to the 
inferred BH mass function.

In the future, we plan to consider other type of clusters such as open and nuclear star clusters which are also believed to be efficient factories of gravitational wave sources \citep{Ziosi2014,2016ApJ...831..187A,2019MNRAS.487.2947D,2019MNRAS.483.1233R,2019MNRAS.486.5008A}. The study of these systems will  require us to add additional physics to \modelbbr .
 
\section{Acknowledgements}
We thank the internal referee of the LVC, Michela Mapelli, for her suggestions, which helped us improve this work. 
MG thanks Duncan Forbes for helpful discussions on the relation between halo mass and GC population mass. We thank the referee for their constructive comments that helped to improved the paper.
FA acknowledges support from a Rutherford fellowship (ST/P00492X/1) from the Science and Technology Facilities Council.
We acknowledge the support of the Supercomputing Wales project, which is part-funded by the European Regional Development Fund (ERDF) via Welsh Government.


%

\end{document}